\documentclass[
 article,
 onecolumn,
 groupedaddress,
 showpacs,
 preprintnumbers,
 amsmath,
 amssymb,
 aps,
 prx,
 prx,
 longbibliography,
 floatfix,
]{revtex4-1}

\usepackage{graphicx}					
\usepackage{verbatim}					
\usepackage{dcolumn}                    
\usepackage[ruled,vlined]{algorithm2e}
\usepackage{color}
\usepackage{bm}                        

\usepackage{orcidlink}                     

\newcommand{\kk}{\kappa}
\newcommand{\sn}{\mathrm{sn}}
\newcommand{\cn}{\mathrm{cn}}
\newcommand{\dn}{\mathrm{dn}}

\newcommand{\const}{\mathrm{const}}

\newcommand{\ds}{\displaystyle}
\newcommand{\dd}{\mathrm{d}}

\newcommand{\J}{\mathrm{J}}
\newcommand{\Tr}{\mathrm{Tr}}
\newcommand{\sgn}{\mathrm{sgn}\,}

\newcommand{\z}{\zeta}
\newcommand{\K}{\mathcal{K}}
\newcommand{\Ks}{\mathcal{K}_\mathrm{s}}
\newcommand{\A}{\mathrm{A}}
\newcommand{\B}{\mathrm{B}}

\newcommand{\eK}{\mathrm{K}}
\newcommand{\eF}{\mathrm{F}}
\newcommand{\eE}{\mathrm{E}}

\newcommand{\pd}{\partial}

\begin{document}

\title{Dynamics of McMillan mappings III. \\
       Symmetric map with mixed nonlinearity}
\author{T.~Zolkin\orcidlink{0000-0002-2274-396X}}
\email{iguanodyn@gmail.com}
\affiliation{Independent Researcher, Atlanta, GA 30341}
\affiliation{(previously at Fermilab, PO Box 500, Batavia, IL 60510-5011)}
\author{S.~Nagaitsev\orcidlink{0000-0001-6088-4854}}
\email{snagaitsev@bnl.gov}
\affiliation{Brookhaven National Laboratory, Upton, NY 11973}
\affiliation{Old Dominion University, Norfolk, VA 23529}
\author{I.~Morozov\orcidlink{0000-0002-1821-7051}}
\email{i.morozov@corp.nstu.ru}
\affiliation{Novosibirsk State Technical University, Novosibirsk 630073, Russia}
\author{S.~Kladov\orcidlink{0000-0002-8005-9373}}
\affiliation{University of Chicago, Chicago, IL 60637}

\date{\today}

\begin{abstract}

This article extends the study of the dynamical properties of the
symmetric McMillan map, emphasizing its utility in understanding
and modeling complex nonlinear systems.
Although the map features six parameters, we demonstrate that only
two are irreducible: the linearized rotation number at the fixed
point and a nonlinear parameter representing the ratio of terms in
the biquadratic invariant.
Through a detailed analysis, we classify regimes of stable motion,
provide exact solutions to the mapping equations, and derive a
canonical set of action-angle variables, offering analytical
expressions for the rotation number and nonlinear tune shift.
We further establish connections between general standard-form
mappings and the symmetric McMillan map, using the area-preserving
H\'enon map and accelerator lattices with thin sextupole magnet as
representative case studies.
Our results show that, despite being a second-order approximation,
the symmetric McMillan map provides a highly accurate depiction of
dynamics across a wide range of system parameters, demonstrating
its practical relevance in both theoretical and applied contexts.
\end{abstract}

\maketitle

\section{\label{sec:Introduction}Introduction}

Among the most extensively studied nonlinear symplectic mappings
of the plane are the area-preserving quadratic H\'enon
map~\cite{henon1969numerical} and the family of transformations
discovered by E. McMillan~\cite{mcmillan1971problem}.
While the H\'enon map serves as a prototype of chaotic dynamics,
the McMillan map represents an integrable
system~\cite{veselov1991integrable}, where the entire phase space
is foliated by constant level sets of its invariant.

The area-preserving quadratic H\'enon map, originally introduced
by Michel H\'enon as a simplified model of the Poincar\'e section
\cite{lichtenberg2013regular} of a dynamical system, has been
used in a variety of fields due to its rich structure.
Serving as a standard model to explore the long-term behavior of
nonlinear systems, it has proven to be a valuable tool for studying
stability, bifurcations, transition from regular to chaotic dynamics
and the associated fractal
structures~\cite{devaney2018introduction,wiggins1990introduction}.
Beyond its significance in mathematical theory, the area-preserving
maps like the H\'enon map have practical applications in physics and
biology, including celestial mechanics, plasma turbulence, nonlinear
optics (light propagation in nonlinear media), and biological and
neural systems (population dynamics and nonlinear network behavior)~\cite{arnold1997celestial,meyer2017introduction,
caldas2018,mori1998,aref1984,ottino1989kinematics,gibbs1985optical,
rivera20101563,may1976,sprott2003chaos,duru2023}.
In essence, while simplified, the H\'enon map captures the
complexity of nonlinear dynamics in area-preserving systems,
making it a powerful tool for modeling conservative dynamical
systems, including Hamiltonian systems where the total energy
is conserved.

On the other hand, the integrable McMillan map is widely
recognized for its generalizations to other measure-preserving
systems, such as the QRT
map~\cite{quispel1988integrable,quispel1989integrable,Nalini2019},
the curve-dependent McMillan mappings discovered by A.~Iatrou and
A.G.~Roberts~\cite{IR2001I,IR2002II,IR2003III}, or the more recent
three-dimensional generalization of QRT maps~\cite{alonso2023three}.
Originally, the McMillan map has been proposed as a toy model
for accelerators, offering a nonlinear but integrable dynamics
without introducing chaotic behavior such as fractal island chains
or complex intersections of homoclinic and heteroclinic orbits.
It has since been generalized to the axially symmetric McMillan
map, describing 2D motion in a 4D phase space, by R. McLachlan,
V. Danilov and E. Perevedentsev~\cite{McLachlan1993,danilov1997two}.
Potential physical realizations, such as electron
lenses~\cite{Danilov1999,Shiltsev_PhysRevSTAB.2.071001}, further
motivate the study of this system (please refer to the second part
of this series and the references cited
within~\cite{zolkin2024MCdynamicsII}).

However, one crucial aspect of integrable systems often overlooked
is their utility as approximations of real-world systems.
Though idealized, integrable models often provide highly accurate
descriptions of physical phenomena.
A historical example is the ancient model of planetary motion, where
the concept of deferent and epicycle predicted planetary positions
from a geocentric perspective.
Although modern celestial mechanics acknowledges the complexity of
the $n$-body problem, the search for accurate approximations persists,
from Fourier series to symplectic Lie algebras and perturbation
theories~\cite{michelotti1995intermediate,bengtsson1997, morozov2017dynamical}.

In particular, Kepler’s laws of planetary motion, which introduced
elliptical orbits, stood out among these methods.
From the perspective of modern Hamiltonian dynamics and dynamical
systems theory, we now understand that it was not merely a
coincidence or a fortunate approximation.
By uncovering the integrable ``core'' of celestial mechanics,
Kepler's laws provided a remarkably precise description of
planetary dynamics and revealed deeper structure within the chaotic
complexity of solar system dynamics.

Similarly, the McMillan map acts as an integrable
approximation~\cite{zolkin2024MCdynamics}
for a broad class of nonlinear mappings in {\it standard form}
~\cite{suris1989integrable}, featuring a {\it typical} force
function (i.e., smooth with at least one nonzero quadratic or
cubic coefficient in its Taylor series), see
Section~\ref{sec:Connection} for details.
This class encompasses many well-known integrable and chaotic
systems, including the H\'enon,
Chirikov~\cite{chirikov1969research,chirikov1979universal},
Cohen~\cite{cohen1993,rychlik1998algebraic},
Brown-Knuth~\cite{brown1983,brown1985,brown1993},
CNR~\cite{cairns2014piecewise,cairns2016conewise}
and recently discovered nonlinear mappings with polygonal
invariants~\cite{ZKN2023PolI,ZKN2024PolII}.

In previous works, we explored the dynamical properties and
applications of the McMillan map, following the foundational
contributions of McMillan~\cite{mcmillan1971problem},
Iatrou, and Roberts~\cite{IR2001I,IR2002II,IR2003III}.
In the first part of this series~\cite{zolkin2024MCdynamics},
we investigated McMillan multipoles, demonstrating that canonical
McMillan octupoles (FO/DO) provide a second-order nonlinear
approximation to the dynamics of an accelerator with a thin
octupole magnet, while the McMillan sextupole (SX) approximates
a lattice with a sextupole magnet to first order.
In the second part~\cite{zolkin2024MCdynamicsII}, we focused on
the axially symmetric McMillan map, deriving a complete set of
canonical action-angle variables and establishing its connection
to a typical standard axially symmetric transformations.

This article concludes our exploration of the most general
symmetric McMillan map, which represents a mixture of two
fundamental nonlinearities: quadratic (sextupole) and cubic
(octupole).
Using the SX-2 sub-family of mappings, which approximate the
quadratic H\'enon map up to second order, we systematically
analyze the system's resonances, including mid-range amplitudes
and boundaries of stability.

Most importantly, we introduce an atlas of intrinsic parameters
that enables compact yet informative mapping of the system’s
behavior.
By establishing correspondences between the integrable McMillan
map and standard nonlinear mappings, we not only link the McMillan
and H\'enon maps but also extend these connections to a broader
class of systems.
This allows us to predict long-term dynamics and evaluate intrinsic
variables such as the rotation number and action variable for a
wide range of nonlinear systems.

\subsection{Article structure}

The article is organized as follows.
In Section~\ref{sec:Intrinsic}, we introduce the general symmetric
McMillan map and analyze the space of its intrinsic parameters. 
Section~\ref{sec:Fixed} provides a comprehensive examination of the
domain and stability of critical points of the invariant, along with
the parameter sets corresponding to degeneracies and singularities.
Section~\ref{sec:Regimes} and \ref{sec:Dynamics} offer an in-depth
analysis and classification of regimes with stable trajectories,
highlighting their analytical properties.
The final Section~\ref{sec:Connection} explores the correspondence
between a generic map in standard form and the symmetric McMillan
map, using the quadratic H\'enon map and an accelerator lattice with
a thin sextupole as examples.
Appendix~\ref{secAp:Action} includes coefficients for the analytical
expressions of the action variable.

\section{\label{sec:Intrinsic}
Form of the map and intrinsic parameters}

Let $\mathrm{M}_\text{SF}:\,\mathbf{Z}\mapsto\mathbf{Z}'$ be an
area-preserving map in the {\it standard form} (SF) from
$\mathbb{R}^2$ to itself, with $\mathbf{Z}=(P,Q)$
\cite{suris1989integrable}:
\begin{equation}
\label{math:SF}
\begin{array}{ll}
\ds \mathrm{M}_\text{SF}:                               &
\ds Q' = P,                                             \\[0.2cm]&
\ds P' =-Q + F(P),
\end{array}
\qquad\qquad\qquad
\begin{array}{ll}
\ds \mathrm{M}_\text{SF}^{-1}:                          &
\ds Q' =-P + F(Q),                                      \\[0.2cm]&
\ds P' = Q.
\end{array}
\end{equation}
Here $(')$ indicates the application of the map, and $F(\cdot)$
is referred to as the {\it force function}.
The most general
{\it symmetric McMillan map}~\cite{mcmillan1971problem,IR2002II}
is then  defined by a special rational function of degree two:
\[
    F_\text{s}(P) = -\frac{\B_0\,P^2 + \mathrm{E}_0\,P + \Xi_0}
                 {\A_0\,P^2 +    \B_0\,P +   \Gamma_0}.
\]
The map is {\it integrable}~\cite{veselov1991integrable}, meaning
that there exists an {\it integral}/{\it invariant of motion}
$\mathrm{K}_\text{s}[P,Q]$:
\begin{equation}
\label{math:KK'}
\forall\,(P,Q)\in\mathbb{R}^2:\,\,
    \mathrm{K}_\text{s}[P',Q'] - \mathrm{K}_\text{s}[P,Q] = 0,    
\end{equation}
that is given by a biquadratic function depending on six parameters
\[
\mathrm{K}_\text{s}[P,Q] =
\begin{bmatrix}
    Q^2 \\ Q \\ 1
\end{bmatrix}^\mathrm{T}
\cdot\left(
\begin{bmatrix}
    \A_0      & \B_0            & \Gamma_0    \\
    \B_0      & \mathrm{E}_0    & \Xi_0       \\
    \Gamma_0  & \Xi_0           & \mathrm{K}_0
\end{bmatrix}
\cdot
\begin{bmatrix}
    P^2 \\ P \\ 1
\end{bmatrix} \right) =
\A_0\,P^2\,Q^2 + \B_0\,(P^2Q+P\,Q^2) + \Gamma_0\,(P^2+Q^2) +
    \mathrm{E}_0\,P\,Q + \Xi_0\,(P+Q) + \mathrm{K}_0.
\]
The transformation in the form (\ref{math:SF}) can be expressed as
the superposition of two anti-area-preserving involutions
$\mathrm{M}_\text{SF} = \mathrm{R}_2\circ\mathrm{R}_1$,
$\mathrm{R}_{1,2} = \mathrm{R}_{1,2}^{-1}$:
\[
\begin{array}{llll}
\ds \mathrm{R}_1:                                       &
\ds Q' = P,                                             &\qquad\qquad\qquad\qquad
\ds \mathrm{R}_2:                                       &
\ds Q' = Q,                                             \\[0.4cm]&
\ds P' = Q,                                             &&
\ds P' =-P + F(Q).
\end{array}
\]
This decomposition arises as a consequence of the map's
invertibility~\cite{lewis1961reversible,devogelaere1950,roberts1992revers},
with each involution preserving the invariant of motion
as described in equation~(\ref{math:KK'}).
The fixed points of the reflections $\mathrm{R}_{1,2}$ define
two fundamental symmetry lines: $l_1$, where $P = Q$, and $l_2$,
where $P = F(Q)/2$.
These lines offer a geometric perspective on the system’s
integrability and help pinpoint the critical points of the
invariant, giving a clearer picture of the map's structure.
Prior to analyzing the dynamics, we remove redundant elements
by identifying intrinsic parameters.

\noindent $\bullet$
There always exists at least one fixed point, which we denote by
$\mathbf{Z}_1$:
$\mathrm{M}_\text{SF}\,\mathbf{Z}_1 = \mathbf{Z}_1$.
We begin by shifting this fixed point to the new origin:
\[
    \mathbf{Z}_1
    \quad\rightarrow\quad
    \overline{\bm{\z}}_1 = (0,0),
\]
using a translation of coordinates:
\[
    \mathbf{Z}
    \quad\rightarrow\quad
    \overline{\bm{\z}}  = (\bar{p},\bar{q})
                        = \mathbf{Z} - \mathbf{Z}_1.
\]
This transformation simplifies the system by ensuring that
$\Xi_0 = 0$.
Next, since adding a constant to the invariant function only
shifts its level sets without altering the equations of motion
or the underlying dynamics, we can redefine the invariant as:
\[
\overline{\mathcal{K}}_\text{s}[\bar{p},\bar{q}] =
\mathrm{K}_\text{s}[\bar{p},\bar{q}] - \overline{\mathrm{K}}_0 =
    \overline{\A}\,\bar{p}^2\,\bar{q}^2 +
    \overline{\B}\,(\bar{p}^2\bar{q}+\bar{p}\,\bar{q}^2) +
    \overline{\Gamma}\,(\bar{p}^2+\bar{q}^2) +
    \overline{\mathrm{E}}\,\bar{p}\,\bar{q}.
\]
Here all barred coefficients
$\overline{\A},\,\overline{\B},\,\ldots,\,\overline{\mathrm{K}}_0$
are obtained from the original invariant expressed in the shifted
variables $\mathrm{K}_\text{s}[\bar{p},\bar{q}]$.
The constant term $\overline{\mathrm{K}}_0$ is then subtracted to
ensure that the newly defined invariant satisfies
$\overline{\mathcal{K}}_\text{s}[0,0]=0$.
This process reduces the number of parameters to four, while
preserving the map's essential dynamics.

\noindent $\bullet$
Next, we rescale the dynamical variables such that
$(\bar{p},\bar{q})=\varepsilon\,(p,q)$.
Dividing the entire invariant by $\varepsilon^2\,\overline{\Gamma}$,
resulting in the form of the symmetric McMillan map
$\mathrm{M}_\text{s}$, which is central to this article:
\begin{equation}
\label{math:Ks}
\Ks[p,q] =
\A\,p^2 q^2 +
\B\,(p^2q + p\,q^2) +
\K_0[p,q],                          \qquad\quad
\K_0[p,q] = p^2 - a\,p\,q + q^2,    \qquad\quad
f_\text{s}(q) =-\frac{\B\,q^2-a\,q}{\A\,q^2+\B\,q+1},
\end{equation}
and where the parameters are as follows:
\[
\Ks = \frac{\overline{\K}_\text{s}}{\varepsilon^2\,\overline{\Gamma}},  \qquad\qquad
\A = \frac{\varepsilon^2\,\overline{\A}}{\overline{\Gamma}},            \qquad\qquad
\B = \frac{\varepsilon  \,\overline{\B}}{\overline{\Gamma}},            \qquad\qquad
a  =-\frac{\overline{\mathrm{E}}}{\overline{\Gamma}}.
\]
The linear part of the invariant, $\K_0$, now depends on a single
intrinsic parameter $a$, which directly relates to the rotation
number at the origin (also known as the {\it unperturbed betatron
tune} in accelerator physics):
\[
    \nu_0 = \frac{\arccos[a/2]}{2\,\pi}.
\]

\noindent $\bullet$
Finally, by selecting an appropriate value for $\varepsilon$, we
can eliminate an additional parameter and arrive at one of two
possible ``normal'' forms of the invariant:
\[
\begin{array}{llll}
\overline{\B} \neq 0:                                       &\qquad
\,\Ks^0\,[p,q] =
\K_0[p,q] + (p^2q + p\,q^2) + \rho\,p^2 q^2,                &\qquad
\ds\varepsilon = \frac{\overline{\Gamma}}{\overline{\B}},   &\qquad
\ds\rho = \frac{\overline{\Gamma}\,\overline{\A}}{\overline{\B}^2},             \\[0.25cm]
\overline{\A} \neq 0:                                       &\qquad
\Ks^\pm  [p,q] =
\K_0[p,q] + r\,(p^2q + p\,q^2) \pm\,p^2 q^2,                &\qquad
\ds\varepsilon = \left|\frac{\overline{\Gamma}}{\overline{\A}}\right|^{1/2},    &\qquad
\ds r = \frac{\overline{\B}\,\sgn(\overline{\Gamma})}
             {|\overline{\Gamma}\,\overline{\A}|^{1/2}}.
\end{array}
\]
In the first form, $\Ks^0\,[p,q]$ reduces to the McMillan sextupole
(SX) limit~\cite{mcmillan1971problem,zolkin2024MCdynamics}, when
$\rho=0$.
The second form, $\Ks^\pm$, corresponds to the focusing/defocusing
McMillan octupoles (FO/DO)
\cite{mcmillan1971problem,IR2002II,zolkin2024MCdynamics},
with $r=0$ and where the sign $\pm$ is determined by
$\sgn\A = \sgn(\overline{\A}/\overline{\Gamma})$.
It's important to note that rescaling the parameters and translating
the dynamical variables preserve the map's standard form, with the
force functions modified as follows:
\[
\begin{array}{ll}
\ds q' = p,                                     &\ds\qquad\qquad\qquad
f_0\,(p) =-\frac{q^2-a\,q}{\rho\,q^2+   q+1},   \\[0.2cm]
\ds p' =-q + f(p),                              &\ds\qquad\qquad\qquad
f_\pm(p) =-\frac{r\,q^2-a\,q}{\pm   q^2+r\,q+1}.
\end{array}
\]

\noindent
{\bf Note}. The two normal forms should be understood as
complementary parametrizations of the same map: the condition
$\rho\rightarrow 0^\pm$ corresponds to $r\rightarrow\infty$ in
$\K_s^\pm$, while  conversely $r \rightarrow 0^+$ for $\K_s^\pm$
corresponds to $\rho\rightarrow\pm\infty$.

\section{\label{sec:Fixed}Fixed points and 2-cycles}

Besides the point at the origin $\bm{\z}_1=(0,0)$, the invariant
$\Ks[p,q]$ can have up to four additional critical
points~\cite{IR2002II}, corresponding to an extra pair of fixed
points located at the intersection of symmetry lines $l_1$ and
$l_2$:
\[
\bm{\z}_{2,3} = \left(\z_{2,3},\z_{2,3}\right):\quad\,\,\,
\z_{2,3} = \frac{-3\,\B \mp \sqrt{\mathcal{R}_1}}
                {4\,\A},
\qquad\quad
\Ks\left[\bm{\z}_{2,3}\right] = \frac{-1}{4\,\A}\,\left[
    (a-2)^2 +
    \frac{(3\,\B)^2}{2\,\A}\,(a-2) +
    \frac{(3\,\B)^3 \pm \mathcal{R}_1^{3/2}}{8\,\A^2}\,\B
\right],
\]
and a 2-cycle defined by the intersection of the second symmetry
line with its inverse, $p=f_\text{s}(q)/2\land q=f_\text{s}(p)/2$:
\[
\bm{\z}^{(2)}_{1,2} =
    \left(\z^{(2)}_{1,2},\z^{(2)}_{2,1}\right):\quad
\z^{(2)}_{1,2} =
    \frac{(a+2)\,\B \mp \sqrt{(a+2)\,\mathcal{R}_2}}
         {2\,\mathcal{R}_0},\qquad
\Ks\left[\bm{\z}^{(2)}_{1,2}\right] =
    \frac{(a+2)^2}{\mathcal{R}_0},
\]
where the superscript $(n)$ is used to distinguish an $n$-cycle,
and
\[
\mathcal{R}_0 = \B^2-4\,\A,
\qquad\qquad\qquad
\mathcal{R}_1 = (3\,\B)^2+8\,(a-2)\,\A,
\qquad\qquad\qquad
\mathcal{R}_2 = (a+10)\,\B^2 - 32\,\A.
\]

\subsection{Domain and singularities}

Nontrivial critical points exist in the real domain when the
expressions under the radicals are positive.
In the parameter space, the corresponding boundaries are:
\[
    B_0^\pm :\,             a = \pm 2,  \qquad\qquad\qquad
    B_1     :\, \mathcal{R}_1 = 0,      \qquad\qquad\qquad
    B_2     :\, \mathcal{R}_2 = 0.
\]
\begin{itemize}
    \item $B_0^+$. One of the points $\bm{\z}_{2,3}$
    (or both if $\B = 0$) merges with the origin, while the
    remaining fixed point and the 2-cycle are given by:
    \[
        \z_2 =-\frac{3\,\B}{2\,\A},\qquad\qquad\qquad
        \z_{1,2}^{(2)} = \frac{-4}{\B \pm \sqrt{3\,\B^2-8\,\A}}.
    \]
    \item $B_0^-$. The 2-cycle merges with the origin,
    with the fixed points being:
    \[
        \z_{2,3} = \frac{-3\,\B \mp \sqrt{(3\,\B)^2-32\,\A}}
                        {4\,\A}.
    \]
    \item $B_1$. This curve separates two domains with
    real and complex values of $\bm{\z}_{2,3}$.
    On this curve, the points coincide $\bm{\z}_2 = \bm{\z}_3$,
    given by:
    \[
        \z_{2,3} = (a-2)\,\frac{2}{3\,\B}.
    \]
    \item $B_2$. The 2-cycle disappears by merging with
    one of the points $\bm{\z}_{2,3}$, with coordinates:
    \[
       -\frac{4}{\B}\,      \left( = \z_{1,2}^{(2)} \right)
        \qquad\qquad\text{and}\qquad\qquad
        \frac{a-2}{a+10}\,\frac{4}{\B}.
    \]
    \item Additionally, there are two curves corresponding to the
    presence of singularities (i.e., a fixed point or a 2-cycle
    moving to infinity):
    \[
    \begin{array}{l}
        S_1:\, \A = 0,      \\[0.25cm]
        S_2:\, \A = \B^2/4 > 0.
    \end{array}
    \]
\end{itemize}

\noindent
For the system on line $S_1$, the invariant $\Ks[p,q]$
transforms to the canonical McMillan sextupole (SX).
In this case, depending on the sign of $\B$, one of the two fixed
points moves to infinity, while the remaining point and 2-cycle are
given by:
\[
    \z_2 = \frac{a-2}{3\,\B},               \qquad\qquad\qquad
    \z_{1,2}^{(2)} = \frac{a+2 \mp \sqrt{(a+2)(a+10)}}{2\,\B}.
\]
For $\A > 0$, there is another line, $S_2$: $\mathcal{R}_0 = 0$,
where one coordinate of the 2-cycle goes to infinity, effectively
making it disappear from the phase space:
\[
    \z_{2,3} = \frac{-3 \mp \sqrt{5+2\,a}}{\B},
    \qquad\qquad
    \z_1^{(2)} =-\frac{2}{\B},
    \qquad\qquad
    |\z_2^{(2)}| = \infty.
\]

\vspace{0.5cm}
\noindent
{\bf Note}.
There are two notions of {\it canonical} McMillan mappings.
First, in McMillan's original sense~\cite{mcmillan1971problem},
he presented four cases of his map, which we have labeled
in~\cite{zolkin2024MCdynamics} as the McMillan sextupole (SX) and
McMillan octupole, with the latter further divided into three
sub-cases: (FO), (DO), and Duffing (DF).
These are canonical both historically and dynamically, as each
represents a fundamental building block of mixed-nonlinearity maps.
Second, in the sense of Iatrou and Roberts~\cite{IR2002II}, canonical
maps refer only to McMillan octupoles, reflecting the fact that most
individual curves of a general McMillan map can be brought to this
form.

\newpage
\subsection{Degeneracy}

\vspace{-0.25cm}
There are cases where different critical points lie on the
same level set of the invariant $\Ks[p,q]$ but occupy different
locations in the phase space.
These situations lead to various types of degeneracies:
\begin{itemize}
    \item $D_2^\pm:\, \B = 0$ for $\A \gtrless 0$.
    The points $\bm{\z}_{2,3}$ become a pair of symmetric fixed
    points, satisfying $\Ks[\bm{\z}_2] = \Ks[\bm{\z}_3]$:
    \[
        \z_{2,3} = \mp\sqrt{\frac{a-2}{2\,\A}},     \qquad\qquad
        \z^{(2)}_{1,2} = \mp\sqrt{\frac{a+2}{-2\,\A}}.
    \]
    In this case, $\Ks[p,q]$ corresponds to the canonical
    McMillan map: focusing (FO) and defocusing (DO) octupoles.
    This degeneracy introduces an additional symmetry to the
    invariant, such that $\Ks[p,q] = \Ks[-q,-p]$,
    reflecting the force function's odd
    character~\cite{roberts1992revers}.
    \item $D_2^*:\, \B^2 = (2-a)\,\A$.
    Along this curve, a symmetric pair is formed involving the
    point at the origin $\bm{\z}_1$ and $\bm{\z}_2$, with the
    coordinates given by:
    \[
        \z_2 = \frac{a-2}{\B},\qquad\qquad
        \z_3 = \frac{\z_2}{2},\qquad\qquad
        \z^{(2)}_{1,2} = \z_3 \mp
        \frac{\sqrt{(a-2)(a+6)}}{2\,\B}.
    \]
    When the origin is shifted to $\bm{\z}_3$, the invariant once
    again becomes the canonical McMillan map, now representing a
    focusing octupole in the Duffing regime (DF), with an unstable
    point at the new origin.
    \item $D_3,\,D_3^*:\, (a+1)\,\B^2 + a^2\A = 0$.
    The last two curves (which are equivalent to each other up to
    a change of parameter $r \rightarrow -r$) describe the scenario
    where one of the fixed points, let say $\bm{\z}_3$, ends up on
    the same level set of the invariant as the 2-cycle
    $\bm{\z}^{(2)}$, forming an isolated 3-cycle:
    \[
    \bm{\z}^{(3)}_1 = \frac{a}{\B}\,(1,1)
    \quad\rightarrow\quad
    \bm{\z}^{(3)}_2 = \frac{a}{\B}\,\left(1,\frac{-2}{a+2}\right)
    \quad\rightarrow\quad
    \bm{\z}^{(3)}_3 = \frac{a}{\B}\,\left(\frac{-2}{a+2},1\right).
    \]
    The remaining fixed point is
    $\ds
    \bm{\z}_2 = \frac{a-2}{a+1}\,\frac{a}{2\,\B}\,(1,1).
    $
    \item $L_{3,4}$.
    Finally, when $D_3$ intersects $S_1$ (i.e., $\A=0$ and $a=-1$)
    or when $D_3$ and $D_3^*$ simultaneously intersect $D_2^-$
    (i.e., $\B = 0$ and $a = 0$), the map becomes periodic.
    In these cases, the system has a linear force function
    $f(q) = a\,q$ and exhibits a rational rotation number of
    $1/3$ or $1/4$, respectively.
    This leads to a state of super degeneracy, where the mapping
    possesses infinitely many integrals of motion, including
    $\K_0[p,q]$, $\Ks[p,q]$ and e.g.polygonal
    structures~\cite{ZKN2023PolI,ZKN2024PolII}.
\end{itemize}

\vspace{-0.45cm}
\subsection{Stability analysis}

\vspace{-0.25cm}
The Jacobian of the transformation $\mathrm{M}_s$ is defined as
\[
\J =
\begin{bmatrix}
    \pd q'/\pd q & \pd q'/\pd p \\[0.4cm]
    \pd p'/\pd q & \pd p'/\pd p
\end{bmatrix} =
\begin{bmatrix}
    0 & 1 \\[0cm]
    -1& -\ds\frac{(\B^2+a\,\A)\,p^2+2\,\B\,p-a}
                 {(\A\,p^2+\B\,p+1)^2}
\end{bmatrix}.
\]
With the help of the expression for the Jacobian trace evaluated
at the fixed point~\cite{IR2002II}:
\[
\tau    (\bm{\z}_*) \equiv 
\Tr\,\J (\bm{\z}_*) =
    \frac{a+4}{\A\,\z_*^2+\B\,\z_*+1} - 4,
\]
for the fixed points $\bm{\z}_{1,2,3}$, we find:
\[
\tau(\bm{\z}_1) = a
\qquad\qquad\text{and}\qquad\qquad
\tau(\bm{\z}_{2,3}) = \frac{1}{2}\,\frac{
    (4-5\,a)\,\B^2 \mp
    (a+4)\,\B\,\sqrt{\mathcal{R}_1} -
    4\,a\,(a-4)\,\A
}{(a+1)\,\B^2+a^2\A}.
\]
Regarding the 2-cycle, its trace is computed as
\[
\tau\left(\bm{\z}^{(2)}\right) \equiv \Tr\,
\left[
    \J\left(\bm{\z}_2^{(2)}\right)\cdot\J\left(\bm{\z}_1^{(2)}\right)
\right] = - 2 +
\frac{\left[
    4\,\A\,\z^{(2)}_1\,\z^{(2)}_2 +
    2\,\B\,\left(\z^{(2)}_1+\z^{(2)}_2\right) -
    a
\right]^2}
{\ds\prod_{\z=\z_{1,2}^{(2)}}\left[\A\,\z^2 + \B\,\z + 1\right]}
\]
simplifying to~\cite{IR2002II}
\[
\tau\left(\bm{\z}^{(2)}\right) = - 2
    - \frac{(a+4)^2(\B^2-4\,\A)}{(a+1)\,\B^2+a^2\A}.
\]

\newpage
\begin{figure}[t!]
    \centering
    \includegraphics[width=\linewidth]{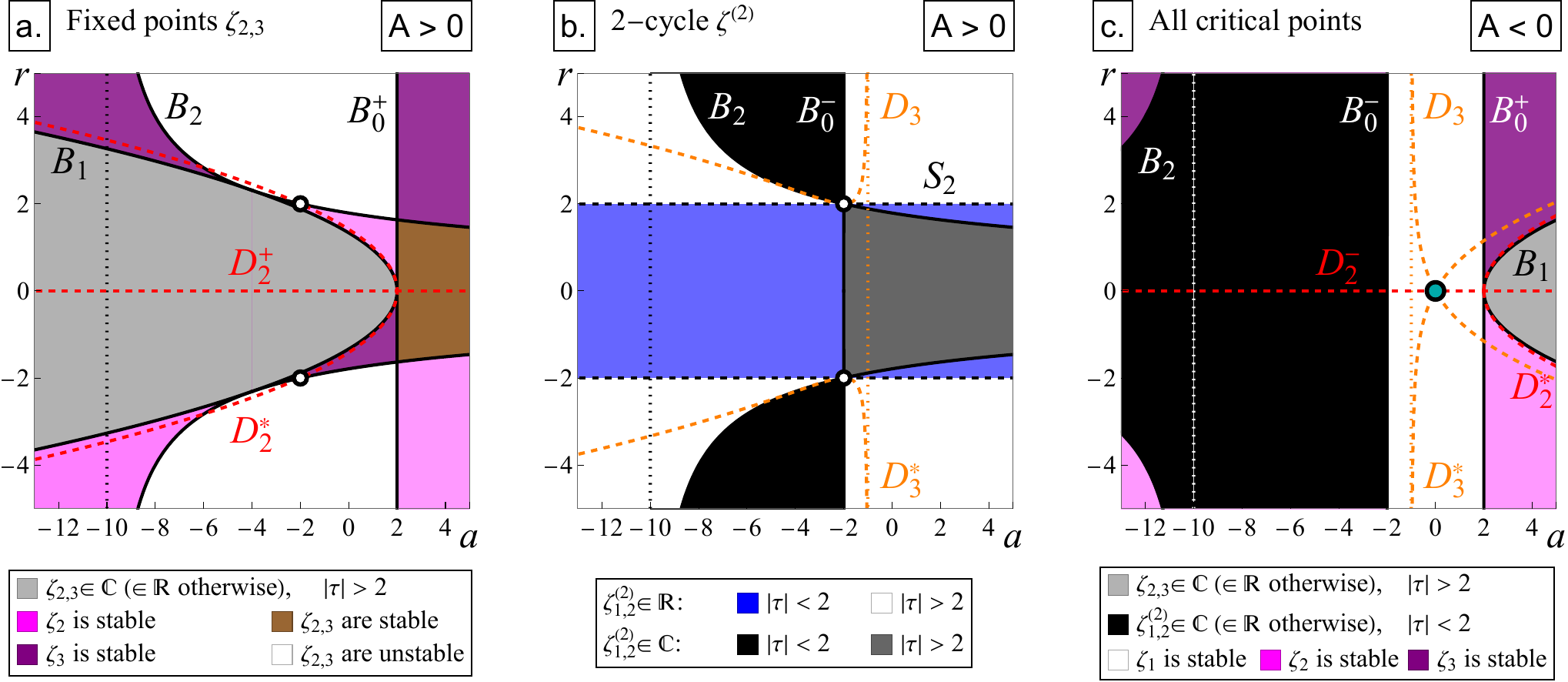}
    \caption{
    Combined diagram illustrating the stability and real/complex
    domains for the fixed points $\bm{\z}_{2,3}$ (subplot a) and
    the 2-cycle $\bm{\z}^{(2)}$ (subplot b) of the symmetric
    invariant $\Ks^+[p,q]$ ($\A > 0$), and for all critical
    points of $\Ks^-[p,q]$ ($\A < 0$) (subplot c).
    Black solid curves represent the boundaries $B_0^\pm$ and
    $B_{1,2}$, while dashed solid lines denote the singularity
    $S_2$, corresponding to $r = \pm 2$.
    Dashed red and orange curves indicate the degeneracies
    $D_2^\pm$, $D_2^*$, $D_3$, and $D_3^*$.
    Additionally, vertical lines shown in dotted black and orange
    represent the asymptotes at $a = -10$ (associated with $B_2$)
    and $a = -1$ (related to $D_3$ and $D_3^*$), respectively.
    }\label{fig:StabBIG}
\end{figure}

The point at the origin, $\bm{\z}_1$, exists for any set of
parameters, and its stability is entirely determined by the
trace of the linearized transformation, requiring $-2<a<2$.
By comparing the absolute value of $\tau$ with 2, we determine
that the stability boundaries for the other critical points are
defined by: $B_0^\pm$, $B_{1,2}$ and $S_2$ for $\A > 0$.
Taking the invariants $\Ks^\pm[p,q]$ as examples, additional
fixed points are real ($\z_{2,3}  \in \mathbb{R}$) and stable
($|\tau| < 2$) under the following conditions:
\[
\begin{array}{ll|lccl}
        &           &\quad a <-10    &\quad -10 < a <-4       &\quad -4 < a < 2        &\quad a > 2             \\[0.1cm]\hline &&&&&\\[-0.3cm]
[\A>0]  & \bm{\z}_2:&\quad r <-r_1^+ &\quad-r_2^+ < r <-r_1^+ &\quad r_1^+ < r < r_2^+ &\quad r < r_2^+         \\[0.25cm]
        & \bm{\z}_3:&\quad r > r_1^+ &\quad r_1^+ < r < r_2^+ &\quad-r_2^+ < r <-r_1^+ &\quad r> -r_2^+         \\[0.25cm]
[\A<0]  & \bm{\z}_2:&\quad r <-r_2^- &\quad -                 &\quad -                 &\quad r < -r_1^-        \\[0.25cm]
        & \bm{\z}_3:&\quad r > r_2^- &\quad -                 &\quad -                 &\quad r >r_1^-
\end{array}
\quad\text{where}\quad
\begin{array}{l}
\ds r_1^\pm = \frac{2\sqrt{2}}{3}\sqrt{\mp(a-2)},   \\[0.25cm]
\ds r_2^\pm = \frac{4\sqrt{2}}{\sqrt{\pm(a+10)}}.
\end{array}
\]
When the 2-cycle is real ($\bm{\z}^{(2)} \in \mathbb{R}^2$),
it can only be stable if $\A > 0$, provided that:
\[
    |r|<2                   \quad       (\text{for }a<-2)
    \qquad\qquad \text{and} \qquad\qquad
    r_2^+<|r|<2             \quad       (\text{for }a>-2),
\]
otherwise, it is unstable.

Fig.~\ref{fig:StabBIG} offers a graphical representation,
illustrating the stability and real domain of critical points
in the parameter space $(a,r)$.
For $\A > 0$, multiple critical points can be stable simultaneously.
Consequently, we present two separate diagrams: one for the fixed
points $\bm{\z}_{2,3}$ (subplot a) and another for the 2-cycle
$\bm{\z}^{(2)}$ (subplot b).
In these diagrams, stability is color-coded: magenta for
$\bm{\z}_2$, purple for $\bm{\z}_3$, blue for the 2-cycle,
and gold for the case where both $\bm{\z}_{2,3}$ are stable.
Areas where the fixed points or 2-cycle fall into the complex
domain with $|\tau|>2$ are depicted in light or dark gray
respectively, and black if $\bm{\z}^{(2)}\in\mathbb{C}^2$
with $|\tau| < 2$.
Regions where both $\bm{\z}_{2,3}$ (subplot a) or $\bm{\z}^{(2)}$
(subplot b) are real but unstable are shown in white.

For $\A < 0$, only one fixed point can be stable for any given
$(a,r)$, and the 2-cycle is always unstable when defined in the
real domain.
Therefore, the stability diagram for all critical points of
$\Ks^-[p,q]$ is combined into a single plot, as shown in
Fig.~\ref{fig:StabBIG}~(c).
In this combined diagram, white, magenta, and purple highlight
stable regions for $\bm{\z}_{1,2,3}$, while black and light gray
denote areas where $\bm{\z}^{(2)}$ and $\bm{\z}_{2,3}$ are complex.

\newpage
\begin{figure}[t!]
    \centering
    \includegraphics[width=\linewidth]{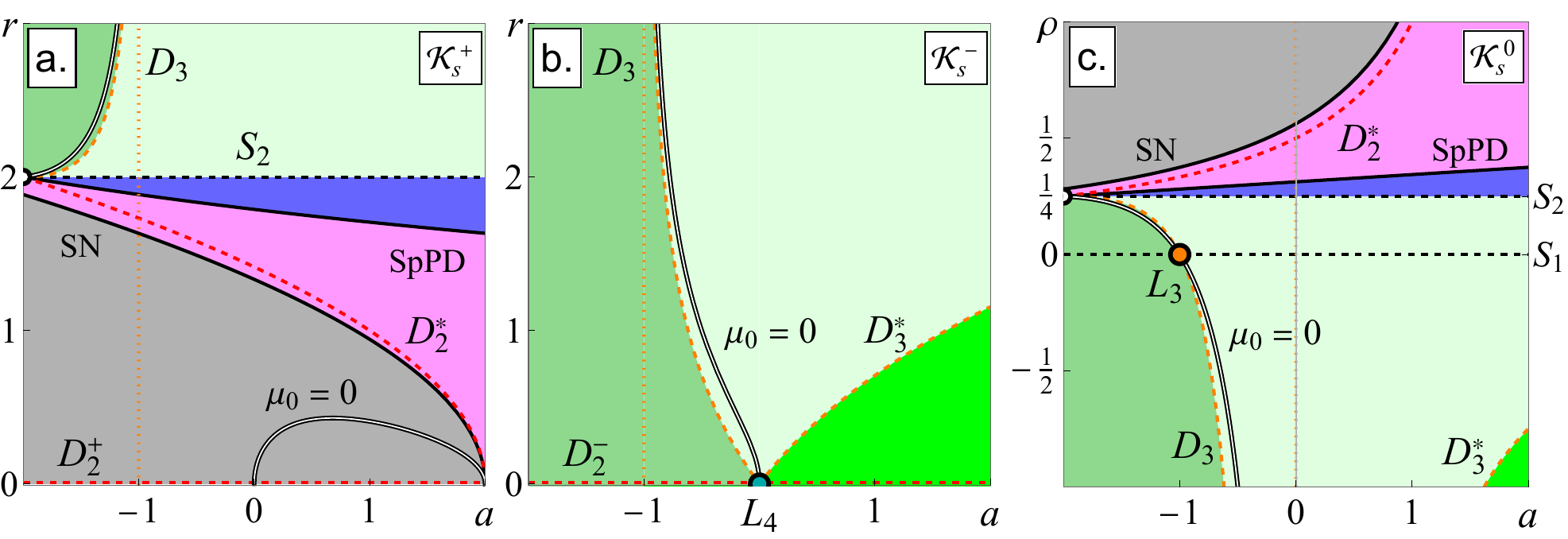}
    \caption{\label{fig:StabSML}
    Atlas depicting stable motion regimes around the origin for
    symmetric invariants $\Ks^\pm[p,q]$ and $\Ks^0[p,q]$.
    The unimodal regime (UM) is shown in gray, the double-well (DW)
    in magenta, the double lemniscate (DL) in blue, and the simply
    connected (SC) regimes in varying shades of green.
    Boundaries of stability ($B$), degeneracy ($D$)
    and singularities ($S$) are color coded according to
    Fig.~\ref{fig:StabBIG}.
    Additional white curves correspond to the set of parameters
    with nonlinear tune shift at the origin equal to zero,
    $\mu_0 = 0$.
    Auxiliary Figs.~\ref{fig:PhSpaceCN} and \ref{fig:PhSpaceSN}
    provide typical phase space diagrams for each regime.
    }
\end{figure}

\section{\label{sec:Regimes}Regimes with stable motion}

\begin{figure}[t!]
    \centering
    \includegraphics[width=\linewidth]{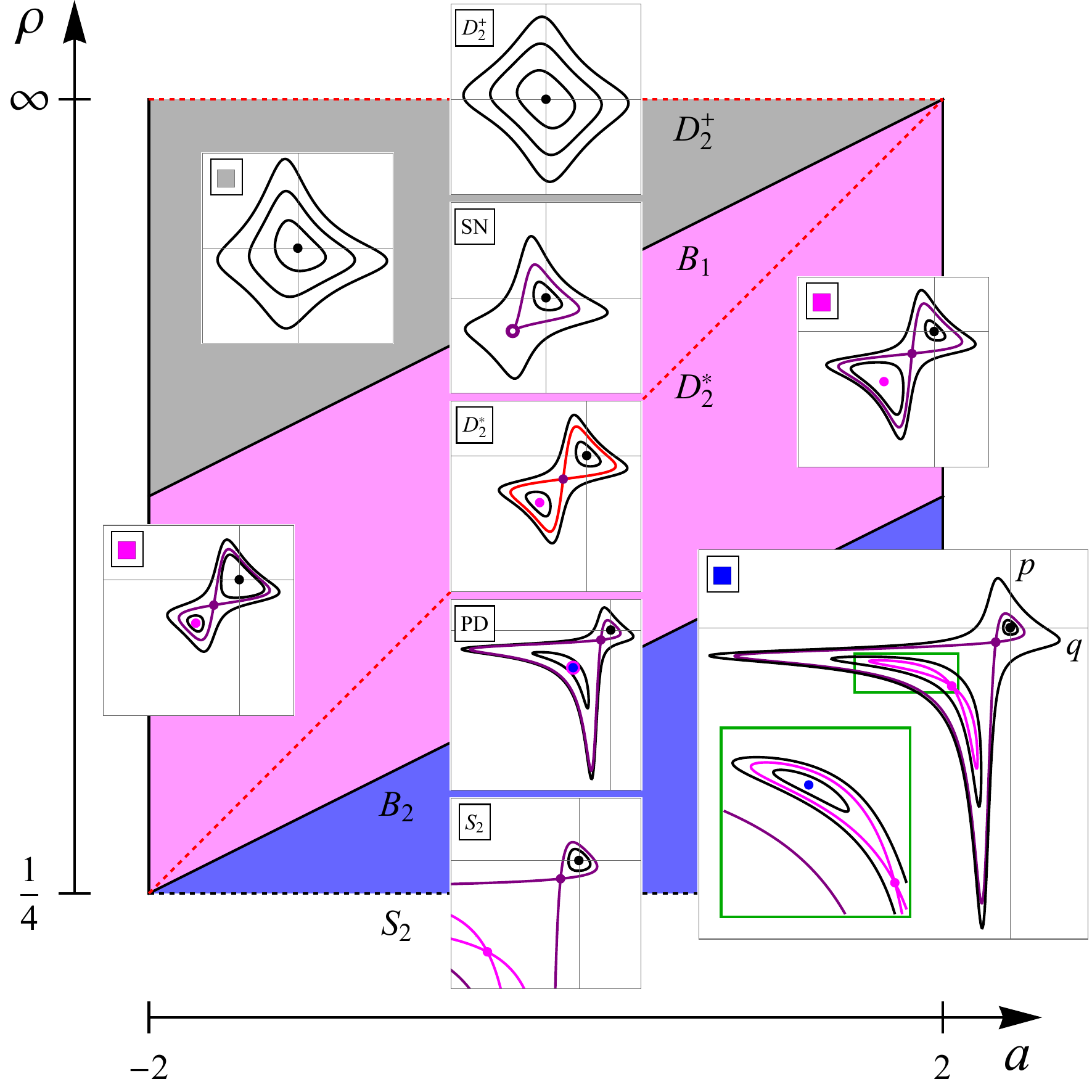}
    \caption{\label{fig:PhSpaceCN}
    Typical phase space diagrams showing stable trajectories around
    the origin for the symmetric McMillan map.
    Isolated fixed points and $n$-cycles, along with their
    corresponding level sets, are highlighted in color, while
    other level sets are depicted in black.
    The plots are schematically arranged in the parameter space
    $(a,\rho)$ for $|a|<2$ and $\rho > 1/4$.
    For more details, refer to Fig.~\ref{fig:StabSML}.
    The plane is delineated by $B_{1,2}$ representing saddle-node (SN)
    and period-doubling (PD) bifurcations, degeneracies $D_2^+$,
    $D_2^*$, and the singularity $S_2$.
    }
\end{figure}

\begin{figure}[t!]
    \centering
    \includegraphics[width=0.99\linewidth]{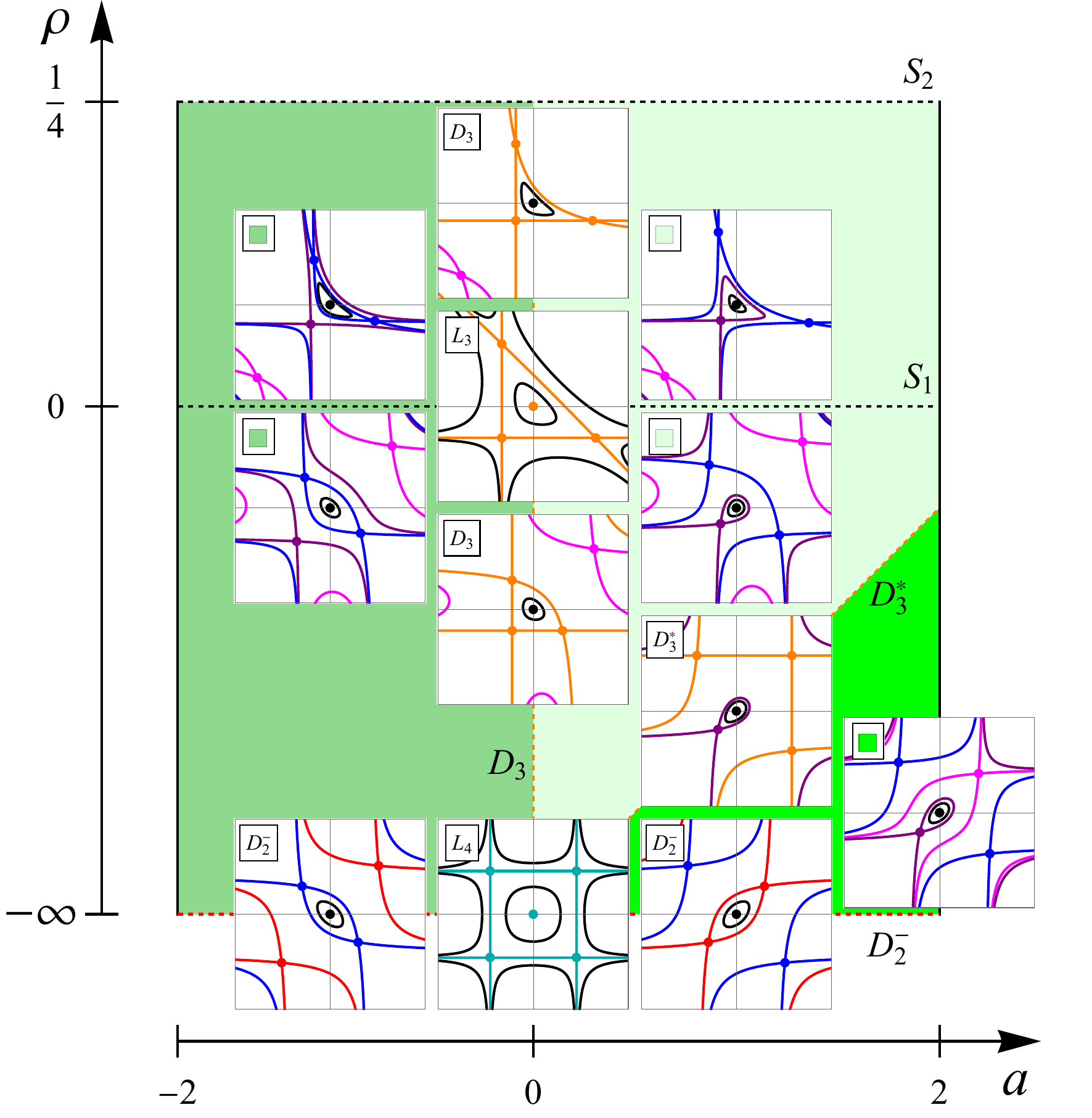}\vspace{-0.25cm}
    \caption{\label{fig:PhSpaceSN}
    Similar to Fig.~\ref{fig:PhSpaceCN}, but illustrating simply
    connected regimes for $|a|<2$ and $\rho < 1/4$.
    The parameter space is outlined by
    degeneracies $D_2^-$, $D_3$, $D_3^*$, and the singularity $S_1$.
    }\vspace{-0.5cm}
\end{figure}

Next, we classify the possible regimes of stable motion, i.e.,
those characterized by bounded and closed level sets of the
invariant, corresponding to topologically distinct configurations
of the phase space.
To ensure the existence of at least one stable fixed point, we can
assume, without loss of generality, that this point is at the origin,
which then requires restricting the parameters to $|a| < 2$;
if this is not the case, we can shift the origin to the stable
point.
Additionally, the diagrams in Fig.~\ref{fig:StabBIG} reveal that
the system allows for the reflection $r \rightarrow -r$, accompanied
by an exchange in the stability properties of the fixed points
$\bm{\z}_{2,3}$.
Consequently, we can further simplify our analysis by focusing on
the case where $r > 0$.
Fig.~\ref{fig:StabSML} presents the reduced parameter space $(a,r)$
for $\Ks^\pm[p,q]$, with color coding indicating different motion
regimes.
Complementary Figs.~\ref{fig:PhSpaceCN} and \ref{fig:PhSpaceSN}
illustrate typical phase space diagrams for all scenarios under
consideration.
Starting with $\Ks^+[p,q]$ in the parameter range $0<r<2$,
we distinguish three qualitatively distinct cases
(Fig.~\ref{fig:PhSpaceCN}):
\begin{itemize}
    \item{Gray area.}
    This region corresponds to the unimodal (UM) regime,
    characterized by a single type of stable motion throughout
    the phase space, and $\bm{\z}_1$ as the only real and stable
    critical point of the invariant.
    As $r$ approaches 0 (line $D_2^+$), the system transitions
    to the focusing McMillan octupole, where oscillations are
    described by the $\cn$ Jacobi elliptic function.
    \item{Magenta area.}
    As $r$ increases and the system crosses $B_1$, it undergoes
    a saddle-node (SN) bifurcation, forming a stable and unstable
    point pair at the cusp of the invariant curve, $\Ks[\bm{\z}_{2,3}]$.
    In this regime, the system exhibits an asymmetric double-well (DW)
    ``potential'' (defined by the characteristic curve), with $\bm{\z}_1$
    representing the minimum on one side of the potential well.
    When parameters lie on $D_2^*$, the potential becomes symmetric,
    and the system behaves as a McMillan octupole in a Duffing regime.
    \item{Blue area.}
    Crossing $B_2$, the fixed point within the second well loses
    stability through a supercritical period-doubling (PD) bifurcation,
    resulting in a stable 2-cycle.
    This creates a regime where a figure-8 separatrix is nested within
    another figure-8 curve, labeled as double lemniscate (DL).
\end{itemize}
Next, for $\Ks^+[p,q]$ with $r>2$ and for $\Ks^-[p,q]$, the system
enters a distinct simply connected (SC) regime, which admits three
sub-cases, see Fig.~\ref{fig:PhSpaceSN}:
\begin{itemize}
    \item{Green-shaded areas.}
    In the SC regime, stable trajectories surround the origin and
    are bounded by a homoclinic or heteroclinic (connecting points
    of 2-, 3-, or 4-cycles) separatrix.
    Apart from the fixed point at the origin, all other isolated
    cycles and fixed points are unstable, and the three green-shaded
    regions differ in the qualitative arrangement of the
    corresponding separatrices in phase space.
\end{itemize}

Fig.~\ref{fig:StabSML}~(c) presents a similar
atlas in the parameter space $(a,\rho)$ for the symmetric
invariant $\Ks^0[p,q]$.
This diagram unifies all possible scenarios of $\Ks^+[p,q]$
(when $\rho > 0$) and $\Ks^-[p,q]$ (when $\rho < 0$) into a
single representation.
Notably, in this depiction, the two lines of degeneracy $D_2^\pm$
(focusing and defocusing octupoles) with $r=0$ correspond to
$\rho = \pm\infty$.
Likewise, the line with singularity $S_1$ (SX), where
$r = \pm\infty$, now corresponds to $\rho = 0$.
Depending on the specific parametrization of the general
biquadratic form, one or a combination of these plots can
be used to describe system behavior.
In the final section, we demonstrate how this diagram serves as
a universal atlas for typical mappings in standard form, using the
quadratic H\'enon map as an example.

By solving for momentum from the expression for the invariant
(\ref{math:Ks})
\[
    p = \frac{1}{2}\,\left(
        f_\text{s}(q) \pm
        \frac{\sqrt{\mathcal{D}_4(q)}}{\A\,q^2+\B\,q+1}
    \right),                \qquad
    \mathcal{D}_4(q) =
        (\B^2-4\,\A)\,q^4 -
        2\,(a+2)\,\B\,q^3 +
        (a^2-4+4\,\A\,\Ks)\,q^2 +
        4\,\B\,\Ks\,q +
        4\,\Ks,
\]
we can classify specific trajectories based on the roots
$q_{1,2,3,4}$ of the characteristic polynomial $\mathcal{D}_4(q)$.

Denoting the roots of the quadratic polynomial in the denominator
of  $p(q)$ as
\[
q_{5,6} = \frac{-\B \mp \sqrt{\mathcal{R}_0}}{2\,\A},
\]
we have the following classification:

\newpage
\begin{enumerate}
\item[$\bullet$] $\mathcal{R}_0 > 0$.
This case corresponds to all simply connected regimes (see Fig.
\ref{fig:PhSpaceSN}), and is referred to as {\it $sn$-like}
trajectories, based on the limiting behavior at $r = 0$ when $\A<0$.
The roots in the denominator are real, $q_{5,6}\in\mathbb{R}$,
and the characteristic polynomial can be factored as
\[
\mathcal{D}_4^{sn}(q) =
    \mathcal{R}_0(q_4-q)(q_3-q)(q-q_2)(q-q_1).
\]
Oscillations around the origin occur between $q_\mp=q_{2,3}$,
such that $q_1<q_2\leq q \leq q_3 < q_4$, and with the roots
$q_{5,6}$ ordered depending on the sign of $\A$ and $B$ as
follows:
\[
[\A<0]:       \,\, q_5 < q_1 < q_4 < q_6,  \qquad\qquad
[\A>0,\,\B>0]:\,\, q_5 < q_6 < q_1,        \qquad\qquad
[\A>0,\,\B<0]:\,\, q_4 < q_5 < q_6.
\]
\item[$\bullet$] $\mathcal{R}_0 < 0$.
In this case, both roots $q_{5,6}\in\mathbb{C}$ become complex
conjugates, leading to the following sub-cases separated by the
curve $B_1$:
\begin{itemize}
    \item[$\circ$] $\mathcal{R}_1 < 0$.
    In the unimodal regime, only two roots of $\mathcal{D}_4(q)$
    are real, $q_{1,2} = q_\mp$, while the other two,
    $q_{3,4} = q_r \mp i\,q_i$, are complex conjugates.
    The characteristic polynomial can be expressed as
    \[
    \mathcal{D}_4^{cn}(q) =
        -\mathcal{R}_0(q_2-q)(q-q_1)[(q-q_r)^2+q_i^2],
    \]
    and these trajectories are referred to as {\it cn-like},
    based on the behavior at $r = 0$, $\A>0$.
    \item[$\circ$] $\mathcal{R}_1 > 0$.
    In the double-well and double lemniscate regimes, we encounter
    two types of stable trajectories.
    When the initial conditions $(\{q_0\},\{p_0\})$ on the invariant
    curve are such that
    \[
        \mathrm{max}\,(0,\K[\bm{\z}_2]) <
        \Ks[\{p_0\},\{q_0\}]           <
        \K[\bm{\z}_3]
    \]
    all four roots of $\mathcal{D}_4(q)$ are real, and these
    trajectories are referred to as {\it dl-} or {\it dr-like}:
    \[
    \begin{array}{l}
    \mathcal{D}_4^{dl}(q) =
        -\mathcal{R}_0(q_4-q)(q_3-q)(q_2-q)(q-q_1),
    \qquad q_1 \leq q \leq q_2 < q_3 < q_4,         \\[0.25cm]
    \mathcal{D}_4^{dr}(q) =
        -\mathcal{R}_0(q_4-q)(q-q_3)(q-q_2)(q-q_1),
    \qquad q_1 < q_2 < q_3 \leq q \leq q_4.
    \end{array}
    \]
    In this notation, the letters {\it l} and {\it r} correspond
    to the left and right ``eyes'' inside the separatrix, respectively,
    with limiting behavior along the curve $D_2^*$ corresponding
    to the Jacobi $\dn$ function.
    Both above $\K[\bm{\z}_3]$, where trajectories round a
    figure-8 pattern, and within the interval
    \[
    \mathrm{min}\,(0,\K[\bm{\z}_2]) <
    \Ks[\{p_0\},\{q_0\}]           <
    \mathrm{max}\,(0,\K[\bm{\z}_2]),
    \]
    at the bottom of the characteristic curve, we once again
    encounter cn-like trajectories.
\end{itemize}
\end{enumerate}

\begin{table}[h]
\centering
\begin{tabular}{lllll}
\hline\hline\\[-0.29cm]
Type of trajectory $\qquad$& $\kappa^2$ & $\Phi(q)$ & $\{q_n\}$ & $\phi_0$                                     \\\hline
sn-like & $\frac{(q_3-q_2)(q_4-q_1)}{(q_3-q_1)(q_4-q_2)}$                                           &
$\arcsin\left[\frac{(q_3-q_1)(q-q_2)}{(q_3-q_2)(q-q_1)}\right]^{1/2}\qquad$                         &
$\frac{q_2(q_3-q_1)-q_1(q_3-q_2)\,\sn^2[\phi_n/2,\kk]}{q_3-q_1-(q_3-q_2)\,\sn^2[\phi_n/2,\kk]}\qquad$&
$\pm2\eF[\Phi(\{q_0\}),\kk]$  \\[0.25cm]&&
$\arcsin\left[\frac{(q_4-q_2)(q_3-q)}{(q_3-q_2)(q_4-q)}\right]^{1/2}$                               &
$\frac{q_3(q_4-q_2)-q_4(q_3-q_2)\,\sn^2[\phi_n/2,\kk]}{q_4-q_2-(q_3-q_2)\,\sn^2[\phi_n/2,\kk]}$     &
$\mp2\eF[\Phi(\{q_0\}),\kk]$  \\[0.25cm]
dl-like & $\frac{(q_2-q_1)(q_4-q_3)}{(q_3-q_1)(q_4-q_2)}$                                           &
$\arcsin\left[\frac{(q_4-q_2)(q-q_1)}{(q_2-q_1)(q_4-q)}\right]^{1/2}$                               &
$\frac{q_1(q_4-q_2)-q_4(q_1-q_2)\,\sn^2[\phi_n/2,\kk]}{q_4-q_2-(q_1-q_2)\,\sn^2[\phi_n/2,\kk]}$     &
$\pm2\eF[\Phi(\{q_0\}),\kk]$  \\[0.25cm]&&
$\arcsin\left[\frac{(q_3-q_1)(q_2-q)}{(q_2-q_1)(q_3-q)}\right]^{1/2}$                               &
$\frac{q_2(q_3-q_1)-q_3(q_2-q_1)\,\sn^2[\phi_n/2,\kk]}{q_3-q_1-(q_2-q_1)\,\sn^2[\phi_n/2,\kk]}$     &
$\mp2\eF[\Phi(\{q_0\}),\kk]$  \\[0.25cm]
dr-like & $\frac{(q_2-q_1)(q_4-q_3)}{(q_3-q_1)(q_4-q_2)}$                                           &
$\arcsin\left[\frac{(q_4-q_2)(q-q_3)}{(q_4-q_3)(q-q_2)}\right]^{1/2}$                               &
$\frac{q_3(q_4-q_2)-q_1(q_4-q_3)\,\sn^2[\phi_n/2,\kk]}{q_4-q_2-(q_4-q_3)\,\sn^2[\phi_n/2,\kk]}$     &
$\pm2\eF[\Phi(\{q_0\}),\kk]$  \\[0.25cm]&&
$\arcsin\left[\frac{(q_3-q_1)(q_4-q)}{(q_4-q_3)(q-q_1)}\right]^{1/2}$                               &
$\frac{q_4(q_3-q_1)-q_1(q_4-q_3)\,\sn^2[\phi_n/2,\kk]}{q_3-q_1-(q_4-q_3)\,\sn^2[\phi_n/2,\kk]}$     &
$\mp2\eF[\Phi(\{q_0\}),\kk]$  \\[0.25cm]
cn-like & $\frac{(q_2-q_1)^2-(u-v)^2}{4\,u\,v}\qquad$                                               &
$\arccos\frac{(q_2-q)v-(q-q_1)u}{(q_2-q)v+(q-q_1)u}$                                                &
$\frac{q_1 u + q_2 v + (q_1 u-q_2 v)\,\cn[\phi_n,\kk]}{u+v+(u-v)\,\cn[\phi_n,\kk]}$                 &
$\pm\eF[\Phi(\{q_0\}),\kk]$   \\[0.25cm]&&
$\arccos\frac{(q-q_1)u-(q_2-q)v}{(q_2-q)v+(q-q_1)u}$                                                &
$\frac{q_1 u + q_2 v - (q_1 u-q_2 v)\,\cn[\phi_n,\kk]}{u+v-(u-v)\,\cn[\phi_n,\kk]}$                 &
$\mp\eF[\Phi(\{q_0\}),\kk]$   \\[0.15cm]\hline\hline
\end{tabular}
    \caption{Elliptic modulus $\kk$, amplitude function $\Phi(q)$,
    and the solution of the map $\{q_n\} = \{p_{n-1}\}$, along with the
    initial phase $\phi_0$, are provided for different types of
    trajectories.
    $\phi_n = 2\,\eK[\kk]\,\{\psi_n\}/\pi = \phi_0+4\,n\,\nu\,\eK[\kk]$
    is the rescaled angle variable such that the sign of $\phi_0$ is
    selected based on the initial conditions $\{p_0\} \gtrless 0$.
    The parameters $u$ and $v$ are defined as
    $u = \sqrt{(q_2 - q_r)^2 + q_i^2}$ and
    $v = \sqrt{(q_1 - q_r)^2 + q_i^2}$, respectively.}
\label{tab:NuPhiQn}
\end{table}

\vspace{-0.3cm}
\section{\label{sec:Dynamics}Dynamical properties}

Using Danilov's Theorem (see Refs.
\cite{zolkin2017rotation,nagaitsev2020betatron,mitchell2021extracting}),
it can be shown that within a simply connected region around the origin,
the map can be expressed in canonical action-angle coordinates
\cite{kolmogorov1954conservation,moser1962invariant,arnol1963small}
\[
\begin{array}{lll}
\ds \mathrm{M}_s:                               &
\ds J' = J,                                     &\qquad\qquad
\ds \{J_n\} = \{J_0\},                          \\[0.4cm]&
\ds \psi' = \psi   + 2\,\pi\,   \nu(J),         &\qquad\qquad
\ds \{\psi_n\} = \{\psi_0\} + 2\,\pi\,n\,\nu(\{J_0\}),
\end{array}
\]
where the rotation number $\nu$ and the action variable $J$ are
defined in terms of integrals involving the characteristic
polynomial $\mathcal{D}_4(q)$:
\[
\nu = \frac{\ds\int_q^{q'}\left(\pd \Ks/\pd p\right)^{-1}\,\dd q}
           {\ds\oint\left(\pd \Ks/\pd p\right)^{-1}\,\dd q}
    = \frac{\ds\int_{q}^{q'} \dd q/\sqrt{\mathcal{D}_4(q)}}
           {\ds 2\,\int_{q_-}^{q_+} \dd q/\sqrt{\mathcal{D}_4(q)}},
    \qquad\qquad
J   = \frac{1}{2\,\pi}\,\oint p\,\dd q
    = \frac{1}{2\,\pi}\,\int_{q_-}^{q_+} \frac{\sqrt{\mathcal{D}_4(q)}}
                                   {\A\,q^2+\B\,q+1}\,\dd q.
\vspace{-0.2cm}
\]
All integrals are evaluated over the constant level set of the
invariant $\Ks[\{p_0\},\{q_0\}] = \const$, with the limits $q_\pm$
being the stop points on a given trajectory, corresponding to two
specific roots of $\mathcal{D}_4(q)$.
The lower bound $q = \{q_0\}$ can be chosen arbitrarily without
affecting the integral, while the upper bound $q' = \{q_0'\}$ is
determined by the mapping equations.

The rotation number $\nu$ is expressed in terms of complete and
incomplete elliptic integrals of the first kind:
\begin{equation}
\label{eq:nu_general}
    \nu = \frac{\eF[\Phi(q'),\kk]}{2\,\eK[\kk]},    
\end{equation}
where the elliptic modulus $\kk$ and the amplitude function
$\Phi(q)$ are detailed in Table \ref{tab:NuPhiQn}.
A particularly convenient choice of the lower bound $q = q_\pm$
allows for:
\[
    q_\pm' = p(q_\pm) = f_\text{s}(q_\pm)/2 =-\frac{1}{2}\,
        \frac{\B\,q_\pm^2-a\,q_\pm}{\A\,q_\pm^2+\B\,q_\pm+1},
\]
which can, if necessary, be expressed in terms of the roots
$q_{1,2,3,4}$, where we use
\[
\A = \frac{h_1}{4}
     \frac{h_1(4\,h_2-h_3^2)-8\,h_0h_3}{h_0(h_1^2 - h_0h_3^2)},
\qquad
\B = \frac{h_1}{h_0},
\qquad
a = \frac{h_1}{4}
     \frac{h_1^3+4\,h_0(2\,h_0h_3-h_1h_2)}{h_0(h_1^2 - h_0h_3^2)},
\]
with
\[
h_0 = \sum_{i=1}^4 q_i,
\qquad
-h_1= \sum_{1 \leq i < j \leq 4} q_i q_j,
\qquad
h_2 = \sum_{1 \leq i < j < k \leq 4} q_i q_j q_k,
\qquad
-h_3= \prod_{i=1}^4 q_i.
\]
Similarly, the action can be calculated analytically as a sum of
five complete elliptic integrals:
\begin{equation}
\label{eq:action_general}
    J = \sqrt{\left|\mathcal{R}_0\right|}\left(c_\mathrm{K} \eK[\kk] +
        c_\mathrm{E} \eE[\kk] +
        c_0 \Pi[\alpha_0, \kk] +
        c_1 \Pi[\alpha_1, \kk] +
        c_2 \Pi[\alpha_2, \kk]\right)/(2\,\A),
\end{equation}
where the modulus $\kk$ matches that in Table~\ref{tab:NuPhiQn},
with the other coefficients provided in Appendix~\ref{secAp:Action}.
To evaluate the nonlinear tune shift $\mu_0 = D_J \nu(0)$ and the
second derivative $D^2_J \nu(0)$, the power series expansion of
$\nu(J)$ can be used:
\[
\begin{array}{l}
\ds 2\,\pi\,(\nu-\nu_0) =
    \frac{s_1}{1!}\,\frac{J}{4-a^2} -
    \frac{s_2}{2!}\,\frac{J^2}{(4-a^2)^{5/2}} +
    \mathcal{O}(J^3),\quad\text{where}              \\[0.35cm]
\ds s_1 = 3\,a\,\A - \frac{(a+1)(a+8)}{2-a}\,\B^2,  \\[0.35cm]
\ds s_2 = a\,(74+7\,a^2)\,\A^2 -
    2\,\frac{208 + 442\,a + 248\,a^2 + 71\,a^3 + 3\,a^4}
            {2-a}\,\A\,\B^2 +
    (a+1)\,\frac{736 + 626\,a + 198\,a^2 + 7\,a^3 - a^4}
                {(2-a)^2}\,\B^4.

\end{array}
\]

Reference~\cite{zolkin2024MCdynamics} provides examples of the
$\nu(J)$ along the lines $r=0$ ($\A=\pm 1$) and $\rho=0$, while
additional illustrations for the set of parameters along the curve
corresponding to the second-order approximation of the H\'enon
quadratic map are provided in the next section.
Finally, the last two columns of Table~\ref{tab:NuPhiQn} offer the
parametrization of the invariant curve, consistent with the results
obtained by methods described in~\cite{IR2002II}.

\newpage
\section{\label{sec:Connection}
Approximating non-integrable maps in standard form}

This section illustrates how the symmetric McMillan map can be
used to approximate chaotic dynamics, exemplified by the quadratic
H\'enon map and accelerator lattices with thin sextupoles.
This connection follows naturally from a more general approach
based on the search for approximate invariants, for a detailed
derivation of which we refer the reader
to~\cite{zolkin2026geometryI-II}.
McMillan mappings play a special role in this perturbative
framework: they provide low-order nonlinear models that capture
essential information on the twist coefficient and detuning, and
their corresponding approximate invariants are associated with an
exact symplectic integrable map.
One important consequence of this approach is its ability to
extend the linear Courant-Snyder formalism to nonlinear regimes,
enabling predictions of amplitude-dependent betatron tune and
the approximate phase-space area occupied by a single particle.
More realistic applications to Fermilab accelerator lattices are
explored in~\cite{zolkin2025geometryIII}.

In~\cite{zolkin2024MCdynamics}, we discuss that for a map in the
standard form:
\[
\begin{array}{l}
\ds q' = p,                                             \\[0.2cm]
\ds p' =-q + f(p),
\end{array}
\]
with a smooth function $f(q)$ and a fixed point at the origin
$f(0)=0$, perturbation theory can be applied to derive an
approximate invariant:
\[
\K^{(n)} = \K_0 + \epsilon\,\K_1 + \epsilon^2\K_2 + \ldots +
    \epsilon^n\K_n:
\qquad
\K^{(n)}[p',q'] - \K^{(n)}[p,q] = \mathcal{O}(\epsilon^{n+1}),
\]
where $\K_m$ are symmetric, homogeneous polynomials of degree
$(m+2)$ in $p$ and $q$.
Here, we introduce a small parameter
$\epsilon>0:\,(p,q)\rightarrow \epsilon\,(p,q)$, deliberately
distinguishing it from the scaling parameter $\varepsilon$.
While $\epsilon$ is used for convenience to separate orders in
perturbation theory and can be set to 1 eventually, the scaling
parameter $\varepsilon$ relates the symmetric biquadratic
$\Ks[p,q]$ to its normal forms $\Ks^0$ and $\Ks^\pm$.

Expanding the force function as a power series in $(\epsilon\,p)$
\[
    f(\epsilon\,p) = a\,\epsilon\,p + b\,\epsilon^2p^2 +
        c\,\epsilon^3p^3 + \ldots,
\]
one can demonstrate that, at the second order in perturbation theory,
the approximate invariant becomes
\[
\K^{(2)}[p,q] = \K_0[p,q] -
    \epsilon\,\frac{b}{a+1}\,(p^2\,q + p\,q^2) +
    \epsilon^2\left[\frac{b^2}{a\,(a+1)} - \frac{c}{a}\right]p^2q^2.
\]
This matches the structure of the symmetric McMillan invariant
$\Ks[p,q]$, where the coefficients are:
\[
\frac{\A}{\epsilon^2} = \frac{b^2}{a\,(a+1)}-\frac{c}{a}, \qquad\qquad
\frac{\B}{\epsilon}   = -\frac{b}{a+1},
\]
and the intrinsic nonlinear parameter characterizing the normal
form $\Ks^0[p,q]$ is:
\[
\rho = \frac{a+1}{a}\left[1-(a+1)\,\frac{c}{b^2}\right].
\]

\subsection{Quadratic H\'enon map}

Here, Fig.~\ref{fig:StabSML} serves as an atlas, revealing the
connection between the symmetric McMillan map and a {\it typical}
mapping in standard form, where either $b \neq 0$ or $c\neq 0$.
As an example, consider the quadratic H\'enon map, characterized
by the force function $f_\text{H\'enon}(p)=a\,p+p^2$.
The corresponding second-order approximate invariant is:
\[
\K^{(2)}_\text{SX-2}[p,q] = \K_0[p,q] -
    \frac{p^2q + p\,q^2}{a+1} +
    \frac{p^2q^2}{a\,(a+1)},
\qquad\qquad
f_\text{SX-2}(p) = \frac{a\,p^2+a^2(a+1)\,p}{p^2-a\,p+a\,(a+1)} =
    a\,p + p^2 + \mathcal{O}(p^4),
\]
with its normal form given by:
\[
\K^{(2n)}_\text{SX-2}[p,q] = \K_0[p,q] + p^2q + p\,q^2 + \rho_n\,p^2q^2,
\qquad\qquad
\rho_n = \frac{a+1}{a}
       = \frac{2\,\cos[2\,\pi\,\nu_0]+1}{2\,\cos[2\,\pi\,\nu_0]},
\]
obtained via the renormalization process described in Section
\ref{sec:Intrinsic} and selecting $\varepsilon=-(a+1)$.
In this section, subscripts and superscripts are used to indicate
system-specific values, such as the force function or fixed points.
For instance, ``H\'enon'' refers to the chaotic H\'enon map,
``SX-2'' refers to the second-order approximate invariant (extending
the first-order SX model) and corresponding McMillan map, and an
additional subscript ``$n$'' is used for its normal form.

\begin{figure}[t!]
    \centering
    \includegraphics[scale=0.63]{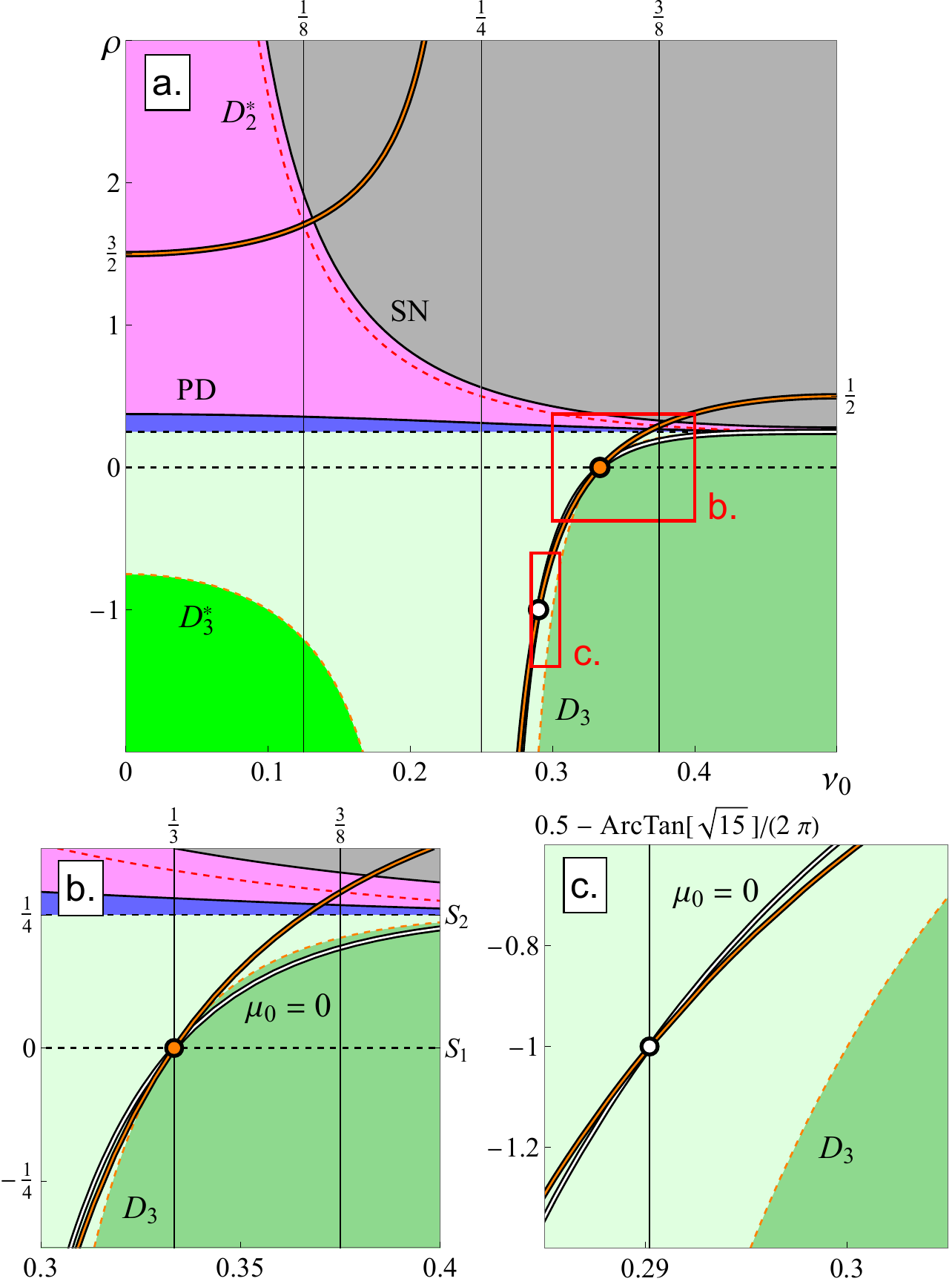}
    \caption{\label{fig:StabilityHNN}
    Atlas illustrating the space of intrinsic parameters for the
    invariant $\Ks^0[p,q]$: the rotation number at the origin
    $\nu_0$ (linear tune) plotted against the nonlinear parameter
    $\rho$.
    The chart is color-coded according to the regimes with stable
    trajectories around the origin (see Fig.~\ref{fig:StabSML}).
    The thick orange curve represents the parameter set $\rho_n(\nu_0)$
    corresponding to the second-order approximate invariant
    $\K^{(2n)}_\text{SX-2}$ for the area-preserving quadratic
    H\'enon map.
    The white curve indicates the parameter set where the nonlinear
    tune shift at the origin is zero, $\mu_0 = 0$.
    The two bottom plots (b and c) provide magnified views of the
    areas outlined in red.
    }\vspace{-0.5cm}
\end{figure}

\newpage
Fig.~\ref{fig:StabilityHNN} depicts a parameter space similar to
the right plot in Fig.~\ref{fig:StabSML}, but this time using
the rotation number at the origin (linear tune) $\nu_0$ instead
of the trace $a$.
The atlas is charted and color-coded in the same manner, with the
thick orange curve representing $\rho_n(\nu_0)$,
establishing a correspondence to the symmetric McMillan maps with
$(\nu_0,\rho)$.
The white curve provides additional information by highlighting
the set of parameters for which the McMillan mapping has zero
tune shift at the origin, $\mu_0=0$.

\newpage
\begin{figure}[t!]
    \centering
    \includegraphics[width=\linewidth]{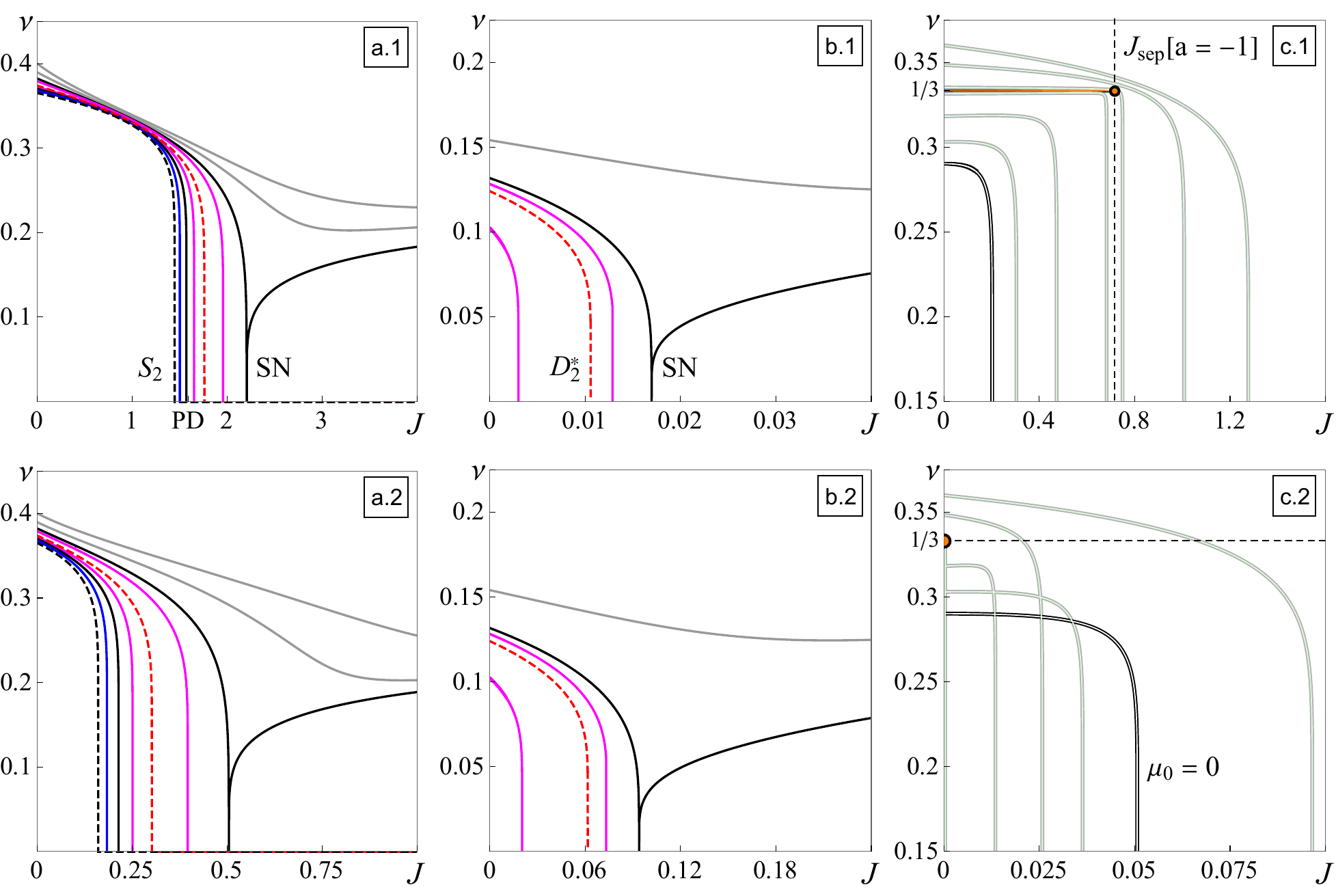}\vspace{-0.2cm}
    \caption{\label{fig:JSamples}
    Rotation number as a function of the action variable $\nu(J)$
    for McMillan mappings with the invariants
    $\K_\text{SX-2}^{(2)}[p,q]$ (bottom row) and its normal form
    $\K_\text{SX-2}^{(2n)}[p,q]$ (top row).
    The columns represent samples from different intervals of the
    linear parameter:
    (a.) below the half-integer $\nu_0 \lesssim 1/2$,
    (b.) just above the integer $\nu_0 \gtrsim 0$, and
    (c.) near the third-order resonance $\nu_0 \approx 1/3$.
    Along the curve $\rho_n(\nu_0)$, sample points
    are selected at key locations:
    bifurcations SN and PD (solid black curves),
    singularity $S_2$ (dashed black curve),
    degeneracy $D_2^*$ (dashed red), and
    super degeneracy $L_3$ (orange).
    Additional samples illustrate typical regimes of motion:
    gray for unimodal (UM), magenta for double-well (DW),
    blue for double lemniscate (DL), and green for simply
    connected (SC).
    The complementary Fig.~\ref{fig:MuSamples} provides detuning at
    the origin, $\mu_0(\nu_0)$, along with the location of the
    typical samples.
    }\vspace{-0.25cm}
\end{figure}

\vspace{-0.2cm}
\subsubsection{Action-angle variables}

\begin{figure}[t!]
    \centering
    \includegraphics[width=0.68\linewidth]{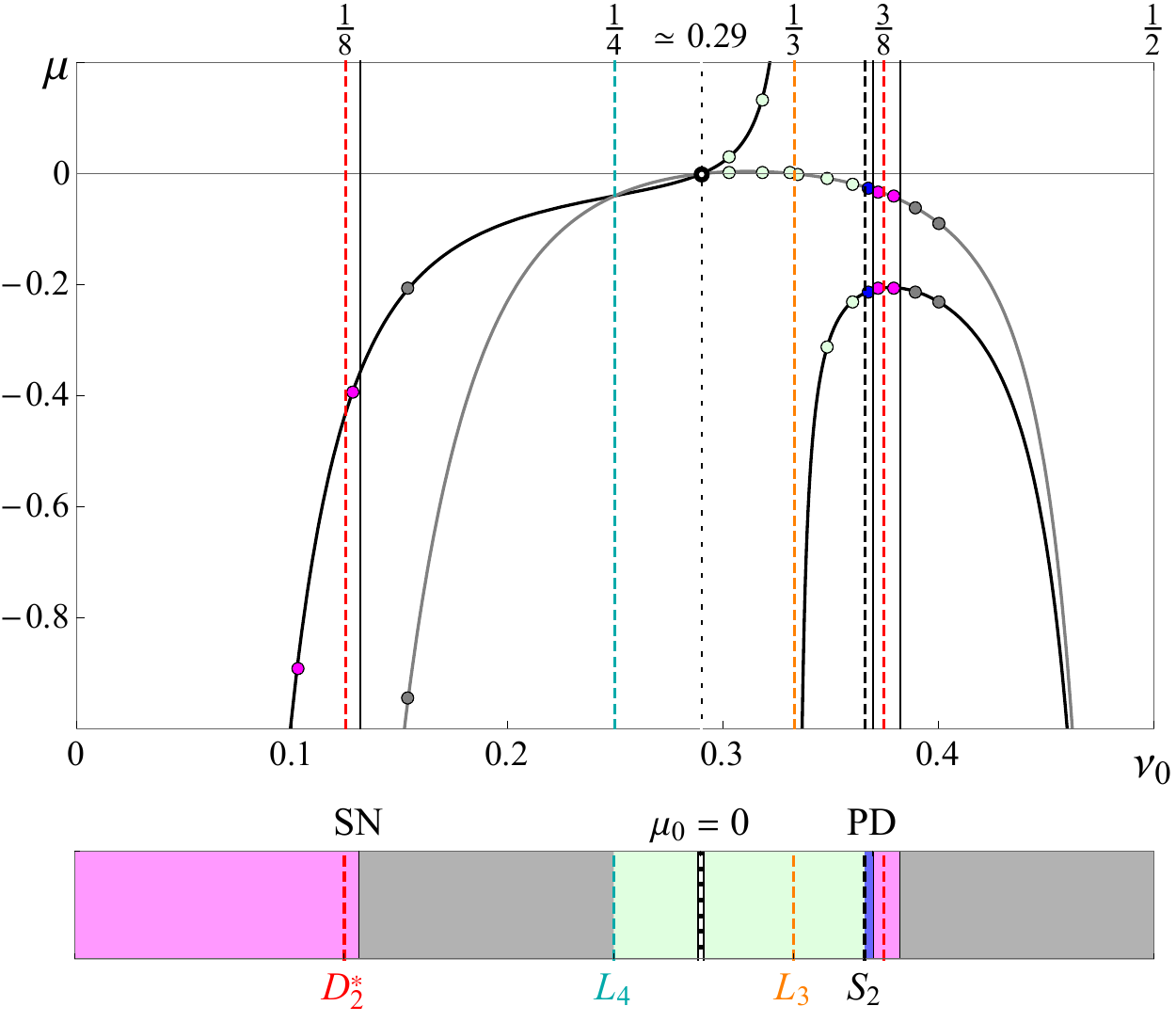}
    \caption{\label{fig:MuSamples}
    Nonlinear tune shift at the origin, $\mu_0(\nu_0)$,
    corresponding to McMillan mappings with the invariants
    $\K_\text{SX-2}^{(2)}[p,q]$ (black curve)
    and its normal form $\K_\text{SX-2}^{(2n)}[p,q]$ (gray curve).
    Colored points are associated with the sample curves shown
    in  Fig.~\ref{fig:JSamples}.
    The legend at the bottom indicates the locations of
    bifurcations (top tick marks), singularities and degeneracies
    (bottom tick marks), and provides color coding for different
    regimes of motion.
    }
\end{figure}

With the approximate invariant in hand, we can now define the
corresponding approximate action variable $J$ and rotation number
$\nu$ for the H\'enon map, using
Eqs.~(\ref{eq:nu_general},\ref{eq:action_general}) and the
appropriate scaling factor provided by $\varepsilon$.
To illustrate the dependence $\nu(J)$, Fig.~\ref{fig:JSamples}
shows samples for McMillan mappings with  the normalized invariant
$\K_\text{SX-2}^{(2n)}[p,q]$ (top row) and approximate invariant
$\K_\text{SX-2}^{(2)}[p,q]$  (bottom row).
The different columns correspond to samples from various intervals
of the linear parameter:
(a.) above the integer resonance $\nu_0\in(0,\arccos[-2/3]/(2\pi))$,
(b.) below the half-integer $\nu_0\in(1/4,1/2)$, and
(c.) in the vicinity of the third-integer resonance $\nu_0 = 1/3$.
At the exact resonance $\nu_0=1/3$, the McMillan map undergoes super
degeneracy, resulting in a linear map.
In case (c.1), a separatrix isolates a simply connected region around
the origin, while in case (c.2), scaling causes this region to vanish
as $J_\mathrm{sep}$ approaches zero (indicated by the orange point).
Additional Fig.~\ref{fig:MuSamples} illustrates the nonlinear
detuning at the origin for both cases.
The black curve represents $\mu_0^\text{SX-2}$, which corresponds
to the H\'enon map and McMillan map SX-2, whereas the gray curve
shows $\mu_0^{(n)}$ for its normal form:
\[
\mu_0^\text{SX-2} = \varepsilon^2 \mu_0^{(n)} = (a+1)^2 \mu_0^{(n)},
\qquad
\mu_0^{(n)} = -\frac{2}{\pi}\,\frac{a+\frac{1}{2}}{(a+1)(a+2)(a-2)^2} =
-\frac{1}{16\,\pi}\,\frac{3\,\cot[\pi\,\nu_0]+\cot[3\,\pi\,\nu_0]}
        {\sin[2\,\pi\,\nu_0]^3}.
\]
The legend at the bottom aligns the dynamical regimes with the
corresponding colors used in Figs.~\ref{fig:StabSML} and
\ref{fig:StabilityHNN}, while the colored points match the
parameters of the sample curves in Fig.~\ref{fig:JSamples}.

\newpage

\subsubsection{Stability diagrams}

In~\cite{zolkin2024MCdynamics}, we demonstrated that although the
McMillan mapping with invariant $\K_\text{SX-2}^{(2)}[p,q]$ is only
a second-order approximation to the quadratic H\'enon map, it
provides an exact expression for the nonlinear detuning at
the origin, $\mu_0^\text{SX-2}$, which aligns with both
numerical simulations and analytical methods like Deprit
perturbation theory~\cite{michelotti1995intermediate} and Lie
algebra treatment~\cite{bengtsson1997, morozov2017dynamical}.
Here, we offer a more systematic analysis of the SX-2 model’s
applicability, especially for large amplitudes.

While the action variable $J$ is useful for infinitesimally small
amplitudes, where the existence of action is ensured by the KAM
theorem~\cite{kolmogorov1954conservation,moser1962invariant,arnol1963small},
larger initial conditions result in the destruction of invariant
tori due to the overlap of nonlinear
resonances~\cite{chirikov1969research,chirikov1979universal}.
To address this, we switch to regular coordinates and compare
the stability areas along the first and second symmetry lines,
$l_{1,2}$, for both the H\'enon map and its integrable
approximation via the symmetric McMillan map, SX-2.

\begin{figure}[t!]
    \includegraphics[width=\linewidth]{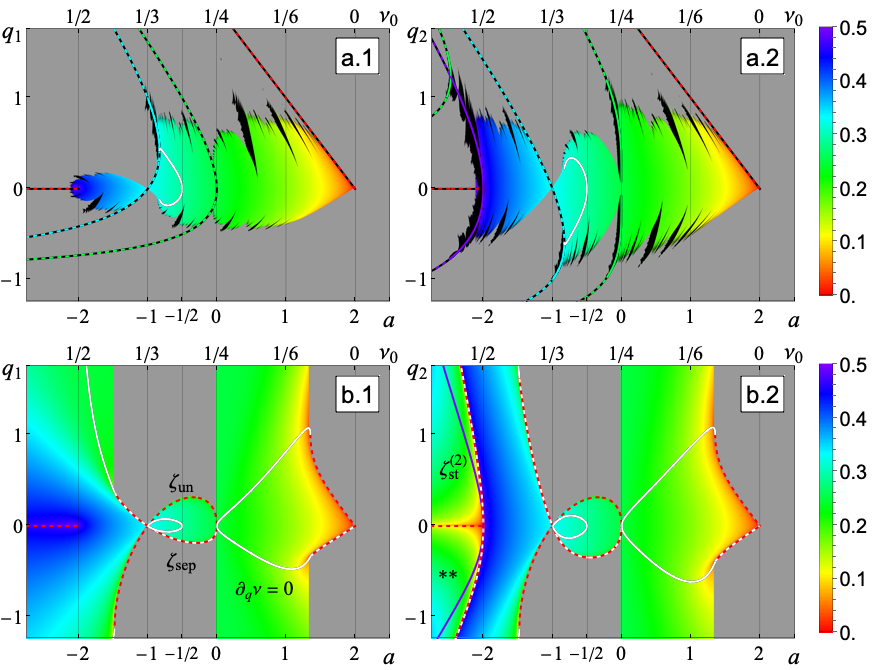}
    \caption{\label{fig:FractalsMcF}
    Stability diagrams for the quadratic H\'enon map (top row a)
    and its integrable approximation via the symmetric McMillan
    map SX-2 (bottom row b).
    Each plot uses a color map to represent the rotation number
    around the origin, $\nu$, as a function of the trace $a\leq 2$
    and the coordinates $q_{1,2}$ along the first (a.1, b.1) and second (a.2, b.2)
    symmetry lines $l_{1,2}$.
    In the top row (a) (H\'enon map), the gray regions indicate unstable
    trajectories escaping to infinity, while black highlights
    mode-locked trajectories within island structures.
    The colored curves represent solutions for isolated fixed points
    (red), 2- (purple), 3- (cyan), and 4-cycles (green);
    the solid indicate stable solutions, and dashed show
    unstable ones.
    In the bottom row (b) (McMillan map SX-2), for $|a|<2$, the rotation
    number is evaluated within the simply connected region around the
    origin.
    In the plot b.2, the area marked with (**)
    corresponds to trajectories encircling 2-cycles inside the
    figure-8 separatrix.
    Since the rotation number in this region is mode-locked to $1/2$,
    the rotation number of the squared map is shown instead.
    Its value along $\bm{\z}^{(2)}_\text{st}$ matches the one obtained
    from the trace of the Jacobian.
    The dashed red and red/white curves indicate the unstable fixed point
    and the corresponding homoclinic orbit (separatrix) along the symmetry
    line, while the solid purple curve represents the stable 2-cycle.
    An additional white curve (all plots) highlights the set of
    coordinates $q \neq 0$ where $\pd_q \nu = 0$.
    The scale at the top represents the rotation number at the origin
    $\nu_0$ for $|a|<2$.
    }
\end{figure}

The top row (a) of Fig.~\ref{fig:FractalsMcF} presents the stability
diagrams for the quadratic H\'enon map.
The color scale indicates the rotation number for initial conditions
along the first (subplot a.1) and second (subplot a.2) symmetry lines,
denoted as $q_{1,2}$.
Gray regions correspond to trajectories that diverge to infinity,
while black indicates mode-locked orbits, i.e., those following
chains of chaotic islands.
The colored curves represent exact analytical solutions for fixed
points and 2-, 3-, and 4-cycles, solid when stable and dashed when
unstable.
The white curve marks the coordinates $q\neq 0$ where $\pd_q \nu = 0$.
For further details on the construction and interpretation of these
stability diagrams, we refer the reader to the dedicated
study~\cite{zolkinHenonSet}.

The bottom row (b) shows similar diagrams for the integrable McMillan
map SX-2.
In this case, for $-2 \leq a \leq 2$, the rotation number is
evaluated only within the simply connected region around the
origin.
For $a<-2$, subplot b.1 shows the rotation number for
trajectories encircling the figure-8 separatrix, while the 
subplot b.2 highlights trajectories inside the ``eyes'' of the figure-8
structure (area of parameters marked with **).
The dashed red/white curve indicates the separatrix crossing along
the appropriate symmetry line.
For $-2 \leq a \leq 2$, this curve corresponds to the homoclinic
orbit attached to $\bm{\z}_\mathrm{un}^\text{SX-2}$, while
for $a<-2$, it represents intersections of symmetry lines with
the figure-8 orbit attached to the unstable point at the origin
(subplot b.2).

Next, we explore the dynamics around the main resonances and
compare the boundaries of stability in each case.

\begin{figure}[t!]
    \centering
    \includegraphics[width=\linewidth]{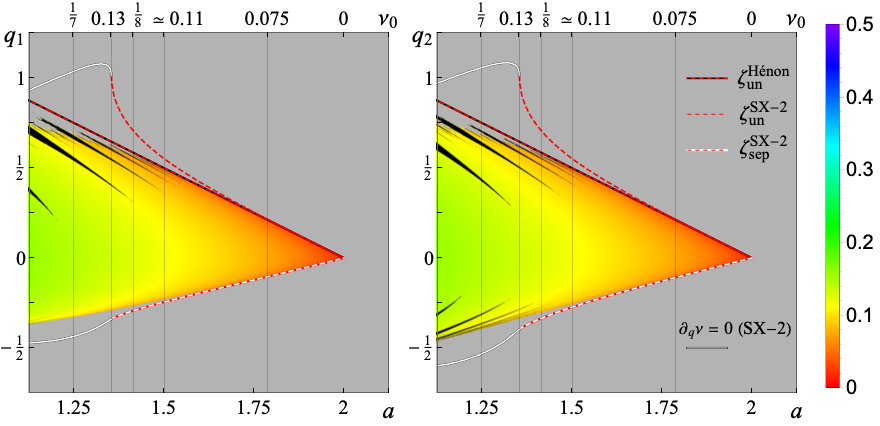}
    \caption{\label{fig:R1SX2}
    Magnification of stability diagrams for H\'enon map above the
    integer resonance (transcritical bifurcation), $\nu_0 > 0$.
    The red dashed curves mark the estimates for the boundaries of
    stability:
    the unstable fixed point $\bm{\z}_\mathrm{un}^\text{H\'enon}$
    for the H\'enon map (red/black) and the unstable fixed point
    from the integrable SX-2 approximation
    $\bm{\z}_\mathrm{un}^\text{SX-2}$ (red).
    The red/white curve corresponds to the intersection of the
    homoclinic orbit (separatrix) with the symmetry line,
    $\bm{\z}_\mathrm{sep}^\text{SX-2}$.
    }
\end{figure}

$\bullet$ {\bf Integer resonance,} $\nu_0 = 0$ ($a=2$){\bf .}
\vspace{0.2cm}

Fig.~\ref{fig:R1SX2} presents a magnified stability diagram for
the H\'enon map near the integer resonance, where $0\leq\nu_0<1/6$.
In addition to showing the coordinates of the unstable fixed point
(dashed red and black)
\[
\z_\mathrm{un}^\text{H\'enon} = 2-a
\]
we include the unstable fixed point for the approximate invariant
(dashed red),
\[
\z_\mathrm{un}^\text{SX-2} =
    \frac{3\,a-\sqrt{a\,(8\,a^2+a-16)}}{4}
\]
as well as the intersection of the symmetry line with the homoclinic
separatrix corresponding to $\bm{\z}_\mathrm{un}^\text{SX-2}$
(shown with dashed red/white).

Introducing the detuning from the integer resonance,
$\delta r_0 = (2-a)$, we find that the deviation between the
unstable fixed points is of third order in $\delta r_0^3$:
\[
\z_\mathrm{un}^\text{H\'enon}-\z_\mathrm{un}^\text{SX-2} =
     -\frac{1}{6}\,\delta r_0^3 + \mathcal{O}(\delta r_0^4).
\]
This indicates that both models yield the same linear estimate
for the slope of the upper stability boundary:
\[
\left.
    \frac{\dd\,\z_\mathrm{un}^\text{H\'enon}}{\dd a}
\right|_{\delta r_0=0} =
\left.
    \frac{\dd\,\z_\mathrm{un}^\text{SX-2}}{\dd a}
\right|_{\delta r_0=0} = -1.
\]
However, the SX-2 model also provides a linear estimate for the
lower boundary:
\[
\left.
    \frac{\dd\,\z_\mathrm{sep}^\text{SX-2}}{\dd a}
\right|_{\delta r_0=0} = \frac{1}{2},
\]
which accurately matches the diagrams in Fig.~\ref{fig:R1SX2}.

Quantitatively, the upper boundary from
$\bm{\z}_\mathrm{un}^\text{SX-2}$ has a relative accuracy within
10\% for $a\in(1.5,2]$ and within 1\% for $a\in(1.78,2]$,
corresponding to $\nu_0\in[0,0.11)$ and $\nu_0\in[0,0.075)$,
respectively.
The lower boundary, determined by the intersection of the symmetry
line with the homoclinic orbit, maintains approximately 10\% accuracy
over the entire applicable range $a\in((3\sqrt{57}-1)/16,2]$, or
$\nu\in[0,0.13)$.

\vspace{0.2cm}
{\it Observation} \#1. The region where SX-2 approximation holds
aligns with the stability diagram's areas that lack significant
mode-locking.
In this area, higher-order resonances minimally overlap, and
stability is governed by the position of the unstable fixed point.

\begin{figure}[t!]
    \centering
    \includegraphics[width=\linewidth]{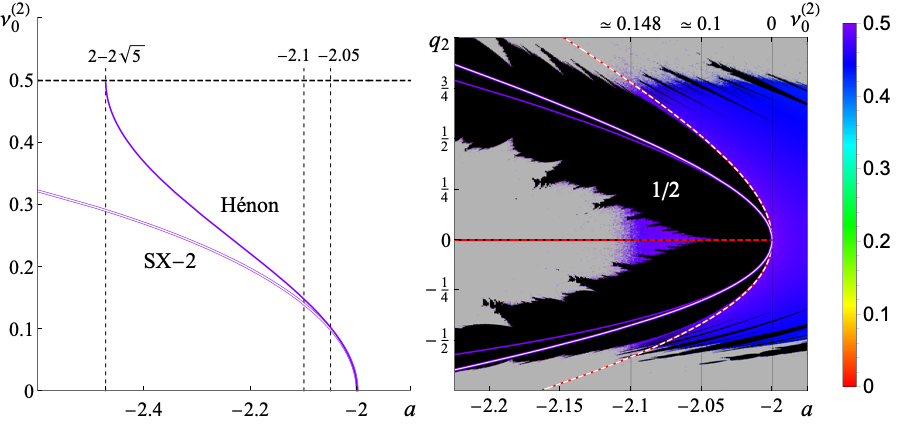}
    \caption{\label{fig:R2SX2}
    The left plot displays the rotation number for the squared map
    evaluated at the stable 2-cycle as a function of the parameter
    $a$: solid purple for the H\'enon map and purple/white for the
    SX-2 symmetric McMillan map.
    The right plot shows a magnified stability diagram for the H\'enon
    map around the half-integer resonance (period doubling bifurcation),
    $\nu_0 = 1/2$, along the second symmetry line.
    The scale at the top corresponds to the rotation number of the
    2-cycle $\bm{\z}^{(2)}_\text{H\'enon}$.
    The red/white dashed curve represents the coordinate of the
    figure-8 separatrix crossing the second symmetry line,
    approximating the outer boundary of the 1/2 mode-locked region
    (indicated by the white label).
    }
\end{figure}

\newpage
$\bullet$ {\bf Half-integer resonance,} $\nu_0 = 1/2$ ($a=-2$){\bf .}
\vspace{0.2cm}

Above the half-integer resonance, ($a<-2$), in both cases, the
fixed point at the origin loses stability through a period-doubling
bifurcation, resulting in the birth of a stable 2-cycle
$\bm{\z}^{(2)} = (\z^{(2)}_{1,2},\z^{(2)}_{2,1})$.
Evaluating the trace of the Jacobian for these 2-cycles gives:
\[
\tau\left( \bm{\z}^{(2)}_\text{H\'enon} \right) =
    14 - a\,(a-4),
\qquad\qquad
\tau\left( \bm{\z}^{(2)}_\text{SX-2} \right) =
    \frac{64+78\,a+24\,a^2+a^3}{a\,(a+1)}.
\]
While the 2-cycle in the SX-2 model is stable for any $a<-2$,
the H\'enon map’s 2-cycle remains stable only in the range
$a\in(2-2\sqrt{5},-2)$.
The left plot in Fig.~\ref{fig:R2SX2} illustrates the rotation
number for these 2-cycles (evaluated from the trace) in purple for
the H\'enon map and purple/white for the SX-2 model.
Although the two models diverge for $a < 2-2\sqrt{5}$, near the
resonance, the difference in the trace values is second-order in
detuning from the resonance:
\[
    \tau\left( \bm{\z}^{(2)}_\text{H\'enon}   \right)-
    \tau\left( \bm{\z}^{(2)}_\text{SX-2}      \right) =
    -4\,\delta r_{1/2}^2 + \mathcal{O}(\delta r_{1/2}^3),
\]
where $\delta r_{1/2} = a + 2$.

To further explore the SX-2 model’s accuracy, we compare the actual
coordinates of the 2-cycles:
\[
\bm{\z}^{(2)}_\text{H\'enon}:\,\,\z^{(2)}_{1,2} =
    \frac{\pm\sqrt{(a+2)(a-6)}-(a+2)}{2},
\qquad
\bm{\z}^{(2)}_\text{SX-2}:\,\,\z^{(2)}_{1,2} =
    \frac{(a+1)\left[
        a\,(a+2) \pm \sqrt{a\,(a+2)(a^2-22\,a-32)}
    \right]}{6\,a+8}.
\]
Both derivatives tend to infinity at $\delta r_{1/2} = 0$:
\[
\left.
    \frac{\dd\,\z^{(2)}_\text{H\'enon}}{\dd a}
\right|_{\delta r_{1/2}=0} =
\left.
    \frac{\dd\,\z^{(2)}_\text{SX-2}}{\dd a}
\right|_{\delta r_{1/2}=0} = \infty.
\]
Thus, we can invert the dependence $\z^{(2)}_{1,2}(a)$ to compare
the {\it terminal} values of parameter $a$
\[
\begin{array}{ll}
\ds a_t^\text{H\'enon} = -q - \frac{4}{2+q} &\ds=
    -2 - \frac{q^2}{2} + \frac{q^3}{4} - \frac{q^4}{8} + \mathcal{O}(q^5),\\[0.25cm]
\ds a_t^\text{SX-2} &\ds=
    -2 - \frac{q^2}{2} + \frac{q^3}{4} + \frac{q^4}{8} + \mathcal{O}(q^5).
\end{array}
\]
Both expansions agree up to $\mathcal{O}(q^4)$.
The right plot in Fig.~\ref{fig:R2SX2} shows a magnified stability
diagram along the second symmetry line, indicating the coordinates
of the 2-cycles for both models (purple curves in the region
mode-locked to $1/2$) and the separatrix crossing
$\z^{(2)}_\mathrm{sep}$ (dashed red/white curve) for the SX-2 model.

\vspace{0.2cm}
{\it Observation} \#2. Similar to the case of integer resonance,
we identify two regions of parameter accuracy: ``high'' and
``medium.''
In the high-accuracy region, where the rotation number for the
2-cycle in the H\'enon map is $\nu^{(2)}_0\in(0,0.1]$, both the
unstable fixed point at the origin and $\z^{(2)}_\mathrm{sep}$
provide good estimates for the mode-locked area.
Further from the resonance, where $\nu^{(2)}_0\in(0.1,0.148]$,
$\z^{(2)}_\mathrm{sep}$ maintains about 15\% accuracy, while
the fixed point at the origin diverges from the boundary of the
mode-locked region.
Beyond $\nu^{(2)}_0 > 0.148$, most invariant tori associated with
orbits around the figure-8 separatrix are destroyed, making
the application of perturbation theory at the origin questionable.

\begin{figure}[b!]
    \centering
    \includegraphics[width=\linewidth]{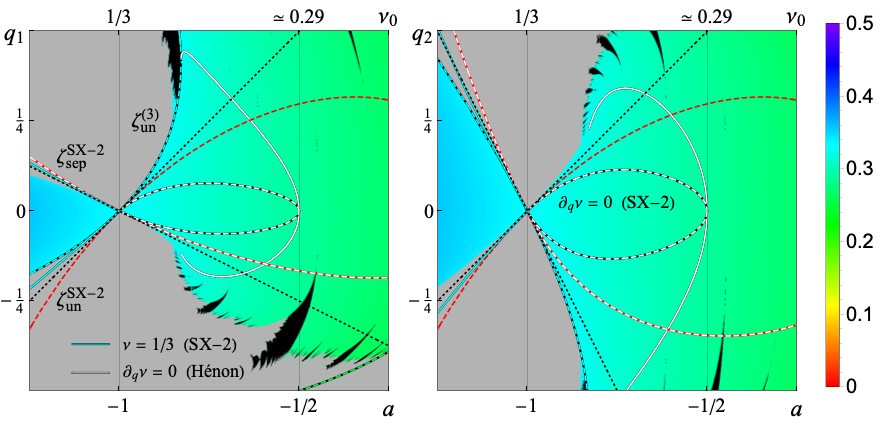}
    \caption{\label{fig:R3SX2}
    Magnification of stability diagrams for the H\'enon map near the
    third-integer resonance (touch-and-go bifurcation),
    $\nu_0 = 1/3$.
    Stability boundaries around the resonance are approximated by
    the unstable 3-cycle $\bm{\z}_\mathrm{un}^{(3)}$ for the
    H\'enon map (dashed cyan), the unstable fixed point (red), and
    the coordinate of the separatrix crossing the symmetry line
    (red/white) in the SX-2 model.
    Their linearized approximation, represented by the derivative
    $\pd_a q_{1,2}(\nu_0 = 1/3)$, is shown with a dotted black line.
    Additional estimates include the coordinates of the non-isolated
    period-3 orbit in the SX-2 model (solid cyan).
    The set of parameters where $\pd_q \nu = 0$  for both the
    H\'enon map and its approximation are distinguished using
    white and dashed white/black curves, respectively.
    }
\end{figure}

\vspace{0.35cm}
$\bullet$ {\bf Third-integer resonance,} $\nu_0 = 1/3$ ($a=-1$){\bf .}
\vspace{0.2cm}

Fig.~\ref{fig:R3SX2} presents a magnified view of the stability
diagrams near the $\nu_0 = 1/3$ resonance.
Since the SX-2 model has an isolated 3-cycle only for $a=-1$, we
propose alternative methods to estimate the stability region for
the H\'enon map.
\begin{enumerate}
    \item Rough estimate:
    The simplest estimate is provided by the fixed point
    $\z_\mathrm{un}^\mathrm{SX-2}$ and the associated separatrix
    crossings, which define a simply connected region around the
    origin, shown with dashed red and red/white curves.
    Although these estimates quickly deviate from the actual
    stability region, they remain accurate up to
    $\mathcal{O}(\delta r_{1/3}^2)$, where $\delta r_{1/3}=a+1$.
    Using this, we can approximate the slopes of the stability
    boundary near the resonance $\nu_0 = 1/3$:
    \[
    \left.
        \frac{\dd\,\z_\mathrm{un}^\text{SX-2}}{\dd a}
    \right|_{\delta r_{1/3}=0} = 1,
    \qquad\qquad
    \left.
        \frac{\dd\,\z_\mathrm{sep}^\text{SX-2}}{\dd a}
    \right|_{\delta r_{1/3}=0} = -2^{2\,s-3},
    \]
    where $s=1,2$ refers to the corresponding symmetry lines.
    \item Improved Estimate:
    A more accurate estimate can be achieved by incorporating the
    non-isolated period-3 orbit in the SX-2 approximation (defined
    for $a<-1$ and shown as a solid cyan curve).
    By using its coordinate we can reduce the error above the
    resonance by approximately 50\%.
    \item Linear Estimates:
    Finally, simple linear estimates obtained from the fixed point
    $\z_\text{un}^\mathrm{SX-2}$ and its corresponding separatrix
    provide a reasonable approximation (shown with black dotted
    lines).
\end{enumerate}

{\it Observation} \#3. Interestingly, both mappings exhibit an
orbit where $\pd_{q}\nu = 0$ which appears for $a<-1/2$: solid
white for the H\'enon map and dashed black/white for the SX-2 model.
In the chaotic case, this structure disappears, giving rise to a
pair of unstable and stable 3-cycles.
In the integrable McMillan SX-2 map, however, it vanishes precisely
at the $\nu_0 = 1/3$.

\begin{figure}[b!]
    \centering
    \includegraphics[width=\linewidth]{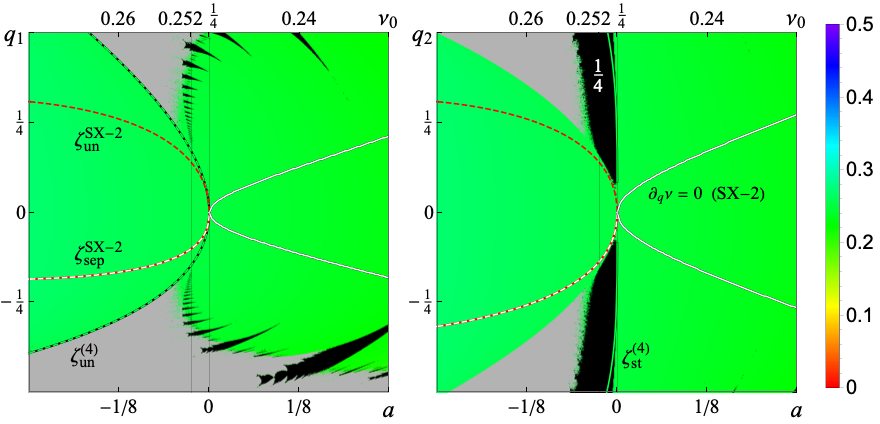}
    \caption{\label{fig:R4SX2}
    Magnification of stability diagrams for the H\'enon map near the
    fourth-integer resonance, $\nu_0 = 1/4$.
    Green curves depict the 4-cycles for the H\'enon map: the unstable 
    $\bm{\z}_\mathrm{un}^{(4)}$ (dashed, left plot) and the stable
    $\bm{\z}_\mathrm{st}^{(4)}$ (solid, right plot), located inside
    the 1/4 mode-locked region (indicated by the white label).
    The red dashed curves show the unstable fixed point and the
    separatrix crossing coordinates (red/white).
    }
\end{figure}

\vspace{0.35cm}
$\bullet$ {\bf Fourth-integer resonance,} $\nu_0 = 1/4$ ($a=0$){\bf .}
\vspace{0.2cm}

Lastly, we examine the stability near the fourth-order resonance 
$\nu_0 = 1/4$, illustrated in Fig.~\ref{fig:R4SX2}.
As in the case of the half-integer resonance, the derivatives of
the stability boundaries' coordinates with respect to the map
parameter $a$ tend to infinity (when $a=0$) in both the SX-2
model and for the 4-cycles in the H\'enon map.
Thus, we once again compare the terminal values $a_t(q)$.
Above the resonance ($\nu_0>1/4$), the first symmetry line $l_1$
intersects with the unstable 4-cycle:
\[
    \z^{(4)}_\text{un} = \frac{-a \pm \sqrt{a\,(a-4)}}{2},
    \qquad\qquad
    a_t =-\frac{q^2}{q+1} = -q^2+q^3+\mathcal{O}(q^4)
\]
which should be compared to the power series expansion of the
coordinates of unstable point $\bm{\z}^\text{SX-2}_\text{un}$
and separatrix crossing $\bm{\z}^\text{SX-2}_\text{sep}$:
\[
a_t^\text{SX-2} = 
     -q^2+\mathcal{O}(q^3).
\]

Along the second symmetry line $l_2$, the $n$-cycle analysis for
the H\'enon map does not reveal any stability boundaries, as the
line intersects only with the stable 4-cycle
$\bm{\z}^{(4)}_\text{st}$, which forms the center of islands
around the origin (mode-locked region marked with a white label).
Above the resonance ($a<0$), stability near the origin
is determined by the stable and unstable manifolds of
$\bm{\z}^{(4)}_\text{un}$, which closely follow the inner
boundary of the mode-locked region for
$\nu_0\in(0.25,\approx 0.252)$.
Below the resonance ($a>0$) the SX-2 model predicts a sharp
transition at $a_t = 0$.
Expanding $a_t$ for the stable 4-cycle
\[
    \z^{(4)}_\text{st} =
    \frac{-a \pm \sqrt{a\,(a-4)+4\sqrt{a\,(a-4)}}}{2},
    \qquad
    a_t =-\frac{q^3+q^2+2-\sqrt{q^4+4\,q^3+4\,q^2+4}}{q\,(q+2)}
        =-\frac{q^4}{4}+\frac{q^6}{8}+\mathcal{O}(q^7),
\]
we observe that the SX-2 estimate remains accurate up to
$\mathcal{O}(q^4)$, as the outer boundary of the mode-locked
region approaches zero faster than $\bm{\z}^{(4)}_\text{st}$.

\vspace{0.2cm}
{\it Observation} \#4. Within the region $\nu_0\in(0.25,\sim 0.252)$,
before the islands separate from the area around the origin, the
homoclinic orbit in the SX-2 model provides a fairly accurate
estimate along the second symmetry line.

\vspace{-0.25cm}
\subsubsection{Mid-range amplitudes}

\begin{figure}[b!]
    \includegraphics[width=0.95\linewidth]{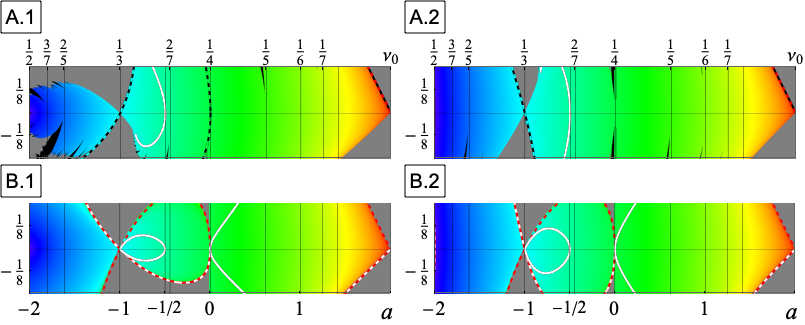}
    \caption{\label{fig:FractalsTiny}
    Magnified view of Fig.~\ref{fig:FractalsMcF} for $a\in[-2;2]$
    ($\nu_0\in[0;1/2]$) and $q_{1,2}\in[-1/4,1/4]$.
    }
\end{figure}

\vspace{-0.25cm}
Before proceeding to the next section, it is important to note
that while the SX-2 model has limited applicability in determining
the precise boundary of stability in the entire range of parameter
$a$, it offers a better fit when considering ``mid-range''
amplitudes, just before the main mode-locking regions.
Although the fractal complexity of the stability region (as seen in
the top row (a) of Fig.~\ref{fig:FractalsMcF}) makes this problem
challenging, the SX-2 approximation for the parameter range 
$q_{1,2}\in[-1/4,1/4]$ (which roughly corresponds to half of the
vertical extent of the stability region) provides a reasonable
estimate for the rotation number $\nu$, as shown in
Fig.~\ref{fig:FractalsTiny}.

Beyond the previously mentioned resonances up to the fourth order,
the main differences in the plots arise primarily due to higher-order
resonances, such as:
\[
\nu_0 =
\frac{1}{5},\frac{2}{5},\frac{1}{6},\frac{1}{7},\frac{2}{7},\frac{3}{7}.
\]
For $a>0$, the white curve in the bottom row (B) of Fig.~\ref{fig:FractalsTiny},
representing an orbit where $\pd_q\nu=0$, marks the upper boundary
of the model's applicability, as the rotation number in the H\'enon
system behaves monotonically with $q_{1,2}$ for this set of parameters
(top row (A)).
\vspace{-0.25cm}

\newpage
\subsection{Accelerator lattice with thin sextupole lens}

The H\'enon map has a wide range of applications, particularly in
modeling particle motion in accelerators~\cite{SYLee4}.
It is especially useful for systems with a nonlinear sextupole magnet,
as it captures the effects of nonlinearity on particle trajectories.
One important application is exploring the {\it dynamic aperture}
--- the region where particle motion remains stable.

In~\cite{zolkin2024MCdynamics}, we show how horizontal motion in
an accelerator lattice, composed of linear (in par-axial approximation)
optical elements (such as drift spaces, dipoles, and quadrupoles)
and a single thin nonlinear lens, can be brought into the standard
form of the map (SF).
The transformations governing the horizontal motion through the
linear part of the lattice are:
\[
    \begin{bmatrix}
      x \\ \dot{x}
    \end{bmatrix}' =
    \begin{bmatrix}
      \cos \Phi + \alpha\,\sin \Phi	& \beta\,\sin \Phi		\\
      -\gamma\,\sin\Phi		& \cos \Phi - \alpha\,\sin \Phi
    \end{bmatrix}
    \begin{bmatrix}
      x \\ \dot{x}
    \end{bmatrix},
\]
followed by the effect of a single thin sextupole lens:
\[
    \begin{bmatrix}
      x \\ \dot{x}
    \end{bmatrix}' =
    \begin{bmatrix}
      x \\ \dot{x}
    \end{bmatrix} -
    \frac{S}{2!}\,
    \begin{bmatrix}
      0 \\ x^2
    \end{bmatrix}.
\]
Here $\alpha(s)$, $\beta(s)$, and $\gamma(s)=(1+\alpha^2)/\beta>0$
are the Courant-Snyder (Twiss) parameters, and $x$ and $\dot{x}$
are the horizontal position and its derivative with respect to the
longitudinal coordinate $s$.
$\Phi$ is the {\it betatron phase advance} related to the {\it bare
betatron tune} $\nu_0$ (i.e., rotation number at the origin) via:
\[
    \Phi = \oint\frac{\dd s}{\beta(s)} = 2\,\pi\,\nu_0,
\]
and $S$ is the integrated sextupole strength:
\[
    S = \int K_x(s)\,\dd s.
\]

While the conventional approach in accelerator physics primarily
describes beam optical functions based on linearized dynamics,
our results extend the Courant-Snyder~\cite{courant1958theory}
formalism to account for nonlinear phenomena up to second order.
Specifically:
\begin{itemize}
    \item Eq.~(\ref{eq:nu_general}) provides an expression for the
    approximated {\it nonlinear betatron tune} $\nu$, where both
    the value of $\nu_0$ and the nonlinear detuning at the origin
    $\mu_0 = \pd_J \nu(0)$ are exact.
    \item Eq.~(\ref{eq:action_general}) for the action variable
    extends the concept of linear single particle {\it emittance}
    and offers an approximation for the phase space area occupied
    by particles.
    \item The approximate invariant
    \[
        \K^{(2)}_\text{SX-2}[p,q] = \K_0[p,q] + \text{higher order terms}
    \]
    extends the linear Courant-Snyder invariant
    (equivalent to $\K_0[p,q]$) and offers an estimate for the
    dynamic aperture by analyzing its critical points and the
    corresponding level sets of the invariant (separatrices).
\end{itemize}

\begin{figure}[b!]
    \includegraphics[width=\linewidth]{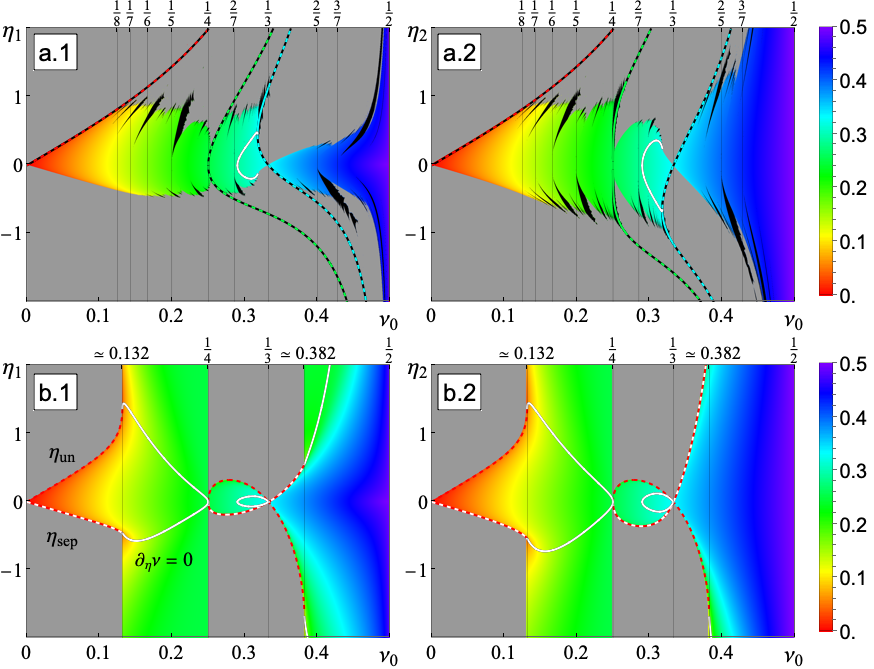}
    \vspace{0.2cm}
    \noindent\makebox[\linewidth]{\rule{\linewidth}{0.4pt}}
    \includegraphics[width=\linewidth]{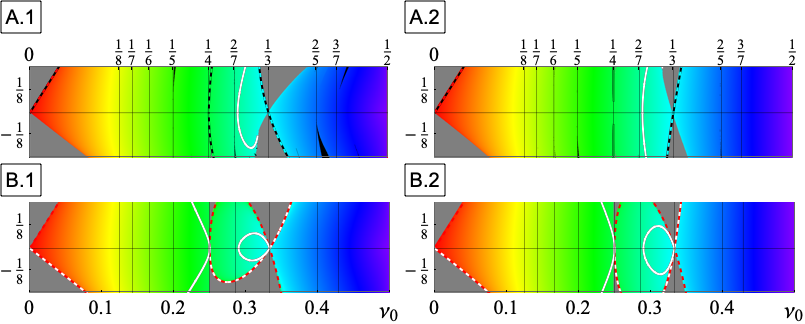}
    \caption{\label{fig:FractalsHnF}
    Stability diagrams for an accelerator lattice with a thin
    sextupole (top row (a)) and its integrable approximation (second
    row (b)) are shown along the first (a.1, b.1) and second (a.2, b.2) symmetry lines.
    The bottom two rows (A and B) provide magnified views of the top plots,
    focusing on the range $\eta_{1,2} \in [-1/4, 1/4]$.
    These plots are equivalent to Figs.~\ref{fig:FractalsMcF} and
    \ref{fig:FractalsTiny} but are expressed in Floquet coordinates 
    $\eta_{1,2}$ along the symmetry lines instead of $q_{1,2}$.
    Additionally, the rotation number at the origin (bare betatron
    tune) $\nu_0$ is used in place of the trace parameter $a$.
    }
\end{figure}

To convert our results, originally obtained in the $(p,q)$ phase
space coordinates, to the accelerator lattice variables, we use
{\it Floquet variables} $(\eta,\dot{\eta})$, representing the
normalized phase space for $(x,\dot{x})$:
\[
\begin{array}{ll}
\ds q/\sqrt{\beta\,\sin\Phi} = \eta,                        &\qquad\qquad\qquad
\ds\eta\,\sqrt{\beta} = x,                        \\[0.25cm]
\ds p/\sqrt{\beta\,\sin\Phi} = \eta\,\cos\Phi + \dot{\eta}\,\sin\Phi, &\qquad\qquad\qquad
\ds\dot{\eta}\,\sqrt{\beta} = \alpha\,x + \beta\,\dot{x}.
\end{array}
\]
Specifically, to rescale the diagrams, we apply the factor
$(\beta\,\sin\Phi)^{-1/2}$.
While the inverse square root of beta-function $\beta$ accounts
for the choice of physical units, the inverse square root of
$\sin\Phi$ introduces significant rescaling, especially as bare
betatron tune $\nu_0$ approaches $1/2$.
Fig.~\ref{fig:FractalsHnF} presents the rescaled versions of
\ref{fig:FractalsMcF} and \ref{fig:FractalsTiny}, where $\eta_{1,2}$
replaces $q_{1,2}$, and $\nu_0$ is used instead of the trace
parameter $a$.

\newpage

\section{Conclusion}

This article presents a comprehensive study of the most general
symmetric McMillan map, emphasizing its role as a universal model
for understanding nonlinear oscillatory systems, particularly
symplectic/area-preserving mappings of the plane in standard form
with typical force functions.
By identifying only two irreducible parameters --- the linearized
rotation number at the fixed point and the coefficient representing
the ratio of nonlinear terms in the biquadratic invariant --- the
McMillan map is shown to be both relatively simple and compact,
yet highly accurate as an integrable approximation for a broad class
of standard-form mappings, especially near main resonances.
Through an in-depth analysis of the map's intrinsic parameters, we
provide a complete solution to the mapping equations and classify
regimes of stable motion.
This general model offers analytical expressions for the nonlinear
tune shift, rotation number, and action-angle variables, and,
also serves as a systematic approach to understanding the
qualitative behavior of nonlinear systems under various parameter
settings.

In the second part of the study, we focus on specific applications
of the symmetric McMillan map to model chaotic systems, specifically
the quadratic H\'enon map and accelerator lattices with thin
sextupole magnet.
By establishing a connection between these systems, we demonstrate
how the McMillan map extends the linear Courant-Snyder formalism,
enabling predictions of dynamic aperture and the nonlinear betatron
tune (rotation number) as a function of amplitude.
We also provide the expression for the approximated single particle
emittance of the beam (the phase space area occupied by particles).
This work underscores the importance of using integrable systems to
accurately model complex nonlinear interactions under certain
conditions, reinforcing the relevance of such models in both
theoretical research and practical applications.

\vspace{0.25cm}
\section{Acknowledgments}

The authors would like to thank Taylor Nchako (Northwestern
University) for carefully reading this manuscript and for her
helpful comments.

\vspace{0.25cm}
\section{Funding}

This manuscript has been authored by Fermi Research Alliance, LLC
under Contract No. DE-AC02-07CH11359 with the U.S. Department of
Energy, Office of Science, Office of High Energy Physics.
Work supported by the U.S. Department of Energy, Office of Science,
Office of Nuclear Physics under contract DE-AC05-06OR23177.
S.K. is grateful to his supervisor, Prof. Young-Kee Kim
(University of Chicago), for her valuable mentorship and continuous
support.

\vspace{0.25cm}
\section{Data abailability}

The data of this paper can be available on request from the
corresponding author.

\vspace{0.25cm}
\section{Author Contribution}
T.Z., S.N., I.M., and S.K. jointly performed the conceptualization,
analytical calculations, and overall planning of the study.
T.Z., I.M., and S.K. wrote the main manuscript text and prepared
the figures.
All authors reviewed the manuscript and contributed equally to this
work.

\vspace{0.25cm}
\section{Conflict of interest}

The authors declare no conflict of interest.

\appendix


\newpage
\section{\label{secAp:Action} Action variable, $J$}

Although the integrands and resulting values of $J$ are real,
the coefficients $c_{1,2}$ and $\alpha_{1,2}$ become complex
when $q_{5,6}\in\mathbb{C}$.
While it should be possible to express the solution entirely in
terms of real-valued functions, to the authors' knowledge, the
form of equation~(\ref{eq:action_general}):
\[
J = \frac{\sqrt{\left|\B^2-4\,\A\right|}}{2\,\A}\left(
        c_\mathrm{K} \eK[\kk] +
        c_\mathrm{E} \eE[\kk] +
        c_0 \Pi[\alpha_0, \kk] +
        c_1 \Pi[\alpha_1, \kk] +
        c_2 \Pi[\alpha_2, \kk]
    \right),
\]
is the most compact and universally applicable for all types of
trajectories.
The coefficients $c_i$ take the following form:
\[
    c_i = \frac{g\,m_i}{\pi},
    \qquad\qquad\qquad
    i=\mathrm{K},\mathrm{E},0,1,2,
\]
where the factor $g$ depends on the type of trajectory:
\[
    g_{sn,dl,dr} = \frac{1}{\sqrt{(q_4-q_2)(q_3-q_1)}}
    \qquad\qquad\text{and}\qquad\qquad
    g_{cn} = \frac{1}{\sqrt{u\,v}}.
\]
The values of $m_i$ and the characteristics $\alpha_i$ for
the elliptic integrals of the third kind are presented in
Table~\ref{tab:Jcoeff}, for the $dl$- and $cn$-like
trajectories.
For $dr$- and $sn$-like trajectories, the corresponding results
can be obtained by the following cyclic substitutions of the roots
from the $dl$-case:
\[
\begin{array}{l}
dr:\qquad
q_1\rightarrow q_3, \qquad
q_2\rightarrow q_4, \qquad
q_3\rightarrow q_1, \qquad
q_4\rightarrow q_2,                             \\[0.2cm]
sn:\qquad
q_1\rightarrow q_2, \qquad
q_2\rightarrow q_3, \qquad
q_3\rightarrow q_4, \qquad
q_4\rightarrow q_1,
\end{array}
\]
and with additional sign changes applied to
$m_{\mathrm{K},0,1,2}\to-m_{\mathrm{K},0,1,2}$ in case of $sn$.

\begin{table}
\centering
\begin{tabular}{lrr}
\hline\hline\\[-0.2cm]
                & $dl$-like trajectory & $cn$-like trajectory \\\hline
                \\[-0.25cm]
$m_\mathrm{K}$  & $(q_4-q_2)(q_4-q_1)$
                & $\ds 2\,\frac{u\,v}{u-v}\,
                \frac{(q_2-q_1)^2\left[|q_5-q_1|^2u-|q_5-q_2|^2v+(u-v)\,u\,v\right]}
                {\left[(q_5-q_1)\,u - (q_5-q_2)\,v\right]
                \left[(q_6-q_1)\,u - (q_6-q_2)\,v\right]}$          \\[0.35cm]
$m_\mathrm{E}$  & $-(q_4-q_2)(q_3-q_1)$
                & $-2\,u\,v$ \\[0.2cm]
$m_0$           & $\ds 2\,(q_4-q_1)\!\left[q_6+q_5-\sum_{i=1}^4 \frac{q_i}{2}\right]$
                & $\ds\frac{u+v}{u-v}\left[q_2^2-q_1^2-2\,(q_2-q_1)\,\Re(q_5) -
                \frac{u^2-v^2}{2}\right]$                           \\[0.35cm]
$m_1$           & $\ds\qquad 2\,(q_4-q_1)\,\frac{(q_5-q_3)(q_5-q_2)}{q_6-q_5}$
                & $\ds\qquad\frac{q_1 u + q_2 v - q_5(u + v)}{q_6-q_5}\,
                \frac{(q_5-q_1)\,u^2-(q_5-q_2)\,v^2+(q_5-q_2)(q_5-q_1)(q_2-q_1)}
                     {(q_5-q_1)\,u - (q_5-q_2)\,v}$                 \\[0.35cm]
$\alpha_0$      & $\ds-\frac{q_2-q_1}{q_4-q_2}$
                & $\ds-\frac{1}{4}\frac{(u-v)^2}{u\,v}$                 \\[0.35cm]
$\alpha_1$      & $\ds-\frac{q_5-q_4}{q_5-q_1}\,\frac{q_2-q_1}{q_4-q_2}$
                & $\ds-\frac{1}{4}\frac{\left[(q_5-q_1)\,u-(q_5-q_2)\,v\right]^2}{(q_5-q_1)(q_5-q_2)\,u\,v}$
                \\[-0.25cm]
                \\\hline\hline
\end{tabular}
\caption{\label{tab:Jcoeff}
    Coefficients $m_{\mathrm{K},\mathrm{E},0,1}$ and parameters
    $\alpha_{0,1}$ used in the action equation~(\ref{eq:action_general})
    for $dl$- and $cn$-like trajectories.
    The values of $m_2$ and $\alpha_2$ can be obtained from $m_1$
    and $\alpha_1$ through the substitution $q_5\leftrightarrow q_6$ in $sn$-like
    case, or by using $m_2 = m_1^*$ and $\alpha_2 = \alpha_1^*$
    for $dl$-, $dr$-, and $cn$-like trajectories.
    Here, ($^*$) denotes the complex conjugate, and $\Re$ represents
    the real part operator.
}
\end{table}


%

\end{document}